\documentclass{astlb}

\usepackage{amsmath}
\usepackage{amsfonts}
\usepackage{amssymb}
\usepackage{amstext}
\usepackage{natbib}
\usepackage{indentfirst}
\usepackage{misccorr}
\bibpunct{(}{)}{;}{a}{,}{,}
\usepackage{lscape}
\usepackage{mathtext}
\usepackage{graphicx}
\usepackage{float}
\usepackage{textcomp}
\usepackage{mathtext}
\usepackage{epsfig}
\usepackage{rotating}
\usepackage{dcolumn}
\usepackage{tabularx}
\usepackage{ulem}
\renewcommand{\deg}{^{\circ}}

\newcommand{\xmm}{\textit{XMM-Newton}}

\newcommand{\beq}{\begin{equation}}
\newcommand{\eeq}{\end{equation}}

\begin{document}
%%%%%%%%%%%%%%%%%%%%
%	Header
%%%%%%%%%%%%%%%%%%%%
\journalinfo{2019}{45}{5}{282}[298]
\date{November 14, 2018}
\submitted{November 14, 2018}
\title{Constraints on the collimated X-ray emission of SS 433 \\ from the reflection on molecular clouds}
\author{\bf 
I.~Khabibullin\email{khabibullin@iki.rssi.ru}\address{1,2},
S.~Sazonov\address{2}
\addresstext{1}{Max-Planck Institute for Astrophysics, Garching, Germany} 
\addresstext{2}{Space Research Institute, Russian Academy of Sciences, Moscow}
%\addresstext{3}{Moscow Institute of Physics and Technology, Dolgoprudny, Russia}
}
\shortauthor{}
\shorttitle{}
%%%%%%%%%%%%%%%%%%%%
%	Abstract
%%%%%%%%%%%%%%%%%%%%
\begin{abstract}
{
We calculate the X-ray signal that should arise due to reflection of the putative collimated X-ray emission of the Galactic supercritical accretor SS 433 on the molecular clouds in its vicinity. The molecular gas distribution in the region of interest has been constructed based on the data of the Boston University-Five College Radio Astronomy Observatory Galactic Ring Survey in $^{13}$CO $J=1\rightarrow0$ emission line, while the collimated emission was assumed to be aligned with the direction of the relativistic jets, which are continuously launched by the system. We consider all the available \textit{Chandra} observations covering the regions possibly containing the reflection signal and put constraints on the apparent face-on luminosity of SS 433 above 4 keV. No signatures of the predicted signal have been found in the analysed regions down to a 4-8 keV surface brightness level of $\sim 10^{-11}$ erg/s/cm$^2$/deg$^2$. This translates into the limit on the apparent face-on 2-10 keV luminosity of SS 433 $L_{X,2-10}\lesssim 8\times10^{38}$ erg/s, provided that the considered clouds do fall inside the illumination cone of the collimated emission. {This, however, might not be the case due to persisting uncertainty in the line-of-sight distances to SS 433 $d_{SS433}$ (4.5-5.5 kpc) and to the considered molecular clouds. For the half-opening angle of the collimation cone larger than or comparable to the amplitude of the jets' precession ($\approx21\deg$), the stringent upper limit quoted above is most relevant if $d_{SS433}<5$ kpc, provided that the kinematic distances to the considered molecular clouds are sufficiently accurate (within $\sim 100$ pc of the adopted values). Dropping the last assumption, a more conservative constraint is $L_{X,2-10}\lesssim 10^{40}$ erg/s for $d_{SS433}=4.65-4.85$ kpc, and yet worse for $d_{SS433}$ outside this range. We conclude that SS 433 is not likely to belong to the brightest ultraluminous X-ray sources if it could be observed face-on, unless its X-ray emission is highly collimated. However, better X-ray coverage of the molecular clouds in the region of interest is needed to eliminate dependence of this conclusion on incidence of the individual clouds inside the putative X-ray illumination cone of SS 433.}
}

\englishkeywords{black holes, accretion, jets, SS 433}
\end{abstract}    
%%%%%%%%%%%%%%%%%%%%
%	Body
%%%%%%%%%%%%%%%%%%%%
%
%%%%%%%%%%%%%%%%%%
\section{1. Introduction}
\label{s:intro}
%%%%%%%%%%%%%%%%%%

%\np
Accretion of matter onto a neutron star or a black hole in a binary system results in an exceptionally high amount of gravitational energy being released in a small region in the close vicinity of the compact object, i.e. at distances of several $R_{in}=3R_{S}=\frac{6GM}{c^2}\sim10^{6} m$ cm, where $m=M/M_{\odot}$ is the mass of the compact object in units of the solar mass $M_{\odot}$, $c$ is the speed of light and $G$ is the gravitational constant. Given that the released energy is efficiently converted into thermal radiation of the infalling matter, the drag force of this radiation starts to strongly affect dynamics of the accretion flow when the accretion rate approaches or exceeds the critical value, the Eddington limit $\dot{M}_{Edd} =3\times 10^{-8}m$ M$_{\odot}$ yr$^{-1}$. As has been envisioned already in the earliest works on accretion theory \citep{1973A&A....24..337S,1980ApJ...242..772A}, and also reproduced in recent numerical simulations \citep{2012ApJ...752...18K,2014SSRv..183..353O}, in this situation, the character of the accretion flow strongly modifies starting from the spherisation radius $R_{sp}\sim \dot{m} R_{in}$, $\dot{m}=\dot{M}/\dot{M}_{Edd} $, inside which it takes the shape of a geometrically and optically thick disc \citep{1973A&A....24..337S}.

%\np
The generic predictions of this picture are, first, that the resulting X-ray emission of the system should get collimated along the axis of the thick disc, and, second, that powerful matter outflows should be launched carrying away significant portions of mass and energy, thereby facilitating self-regulation of the process \citep{1973A&A....24..337S}. Although this implies that the feeding rate of the innermost regions will stay close to the critical value, so that the total luminosity of the source cannot exceed the Eddington limit by a large factor \citep[e.g.][]{2007MNRAS.377.1187P}, strong collimation of this emission might still result in a very high \textit{apparent} luminosity of the source, if observed along the disc axis \citep[e.g.][]{2001ApJ...552L.109K}. 

%\np
It is currently believed that this situation should indeed take place in the case of so-called ultraluminous X-ray sources (ULXs), off-nuclear extragalactic X-ray sources with the apparent X-ray luminosity exceeding the Eddington limit for a 10 $M_{\odot}$ black hole, i.e. $ \approx 1.4\times 10^{39}$ erg/s, with some extreme cases where the apparent luminosity exceeds $10^{41}$ erg/s (see \cite{2017ARA&A..55..303K} for a review). Although some of these sources have proved to be powered by accretion onto neutron stars \citep{2014Natur.514..202B,2016ApJ...831L..14F,2017MNRAS.466L..48I,2018MNRAS.476L..45C}, and some can even harbour a black hole of intermediate ($10^3-10^5 M_{\odot}$) mass \citep[see][and references therein]{2018arXiv180207149C}, the model of supercritically accreting stellar mass black holes seems to fit well most of the available observational data, both in terms of characteristics of individual sources \cite[e.g.][]{2014ApJ...793...21W,2015NatPh..11..551F} and properties of their population \citep{2001ApJ...552L.109K,2004NuPhS.132..369G,2005MNRAS.356..401R,2012MNRAS.419.2095M,2017MNRAS.466.1019S}. The actual accretion rates and degrees of collimation are, however, still uncertain, leaving open also the question about their total (i.e. angular-integrated) luminosity. The latter is of particular importance because ULXs dominate in the total X-ray output of normal star-forming galaxies in the local Universe  \citep{2012MNRAS.419.2095M,2017MNRAS.466.1019S}. Extrapolation of the observed properties of the ULX population suggests that their analogues at redshift $z\sim10$ could play a significant role in the early heating of the Universe before the epoch of reionization \citep[see e.g.][and references therein]{2017AstL...43..211S}.  

%\np
If there were bona fide ULXs in our Galaxy, it would be possible to perform in-depth studies of their actual accretion picture and impact on the surrounding medium, but, unfortunately, none are known. However, it has been speculated for a long time \citep{Fabrika2001,2006MNRAS.370..399B,2007MNRAS.377.1187P} that the very peculiar Galactic microquasar SS 433 might belong to this class, because the mass supply rate provided by the donor star is estimated to persistently  exceed the critical value for any reasonable mass of the relativistic compact object \citep{2018MNRAS.479.4844C} by at least two orders of magnitude \citep{2004ASPRv..12....1F}. The fact that the \textit{apparent} X-ray luminosity of SS 433 is actually quite low, $ L_{X}\sim 10^{36}$ erg/s \citep{1996PASJ...48..619K,2002ApJ...564..941M,2005A&A...431..575B}, i.e. three orders of magnitude dimmer than for canonical ULXs, has been commonly attributed to the nearly edge-on (inclination $i=78\deg$, \cite{2004ASPRv..12....1F}) orientation of the binary system with respect to the line of sight, so that the emission from the innermost disc regions is obscured by the geometrically thick outer parts of the disc and by a wind outflowing from it. As a result, the bulk of the source's X-ray luminosity is in fact provided by a pair of mildly-relativistic baryonic jets and, possibly, some reverberations of the central engine's emission. Functioning of the powerful central engine is evidenced by (i) the very high kinetic luminosity carried by the jets, $\sim 10^{39}$ erg/s \citep[e.g.][]{2016MNRAS.455.1414K,2018AstL...44..390M}, (ii) possible reverberations of its obscured emission in the hard X-ray part of the observed spectra \citep{2010MNRAS.402..479M}, and (iii) the high UV luminosity of the source \citep{1997A&A...327..648D}. 

%\np
For observers seeing SS 433 face-on, it would appear like a canonical ULX or an ultraluminous supersoft source (ULS, \citealt{2016MNRAS.456.1859U,2016MNRAS.457.3963K}). Assuming that its ULX-like emission is collimated along the accretion disc axis, which is, plausibly, aligned with the direction of the relativistic jets, one can easily predict the geometry of the region illuminated by this emission. Taking into account the remarkable stability of the system over more than 40 years of available observations \citep[e.g.][]{2018ARep...62..747C} and the very high penetrating power of X-rays above 3 keV, the extent of this illumination region can easily exceed tens of parsecs. Moreover, SS 433 is located inside the radio nebula W 50, which is believed to be a $\sim 10^4$ years old supernova remnant severely deformed and revitalized by the relativistic jets \citep{2011MNRAS.414.2838G} . This hints that the source's activity period has lasted thousands of years, implying that the illumination region might be $\sim$ kpc long. Thanks to the very precise determination of the jets' 3D direction and $\sim 10$\% accuracy of the distance estimate to the system, the predicted illumination region can be put into the context of SS 433's Galactic environment in order to check how the latter is affected by the ULX-like emission from SS 433. 

%\np
Namely, one might expect a fraction of the hard ($\gtrsim 3$ keV) X-ray collimated radiation of SS 433 to be reflected by atomic and molecular gas, ubiquitously distributed in the plane of the Galaxy. This effect has been thoroughly studied, both theoretically and observationally, particularly in application to the reflection of past Sgr A* flares on molecular clouds in the Galactic center region \citep[e.g.][]{2017MNRAS.471.3293C} and the contribution of scattered X-ray emission of X-ray binaries to the Galactic Ridge X-ray emission \citep{2014A&A...564A.107M}. In the optically thin regime (holding up to the scatterer's column density $N_{H}\sim 10^{23}$ cm$^{-2}$), the reflection signal above 3 keV scales linearly with the incident flux and the total mass of the reflecting gas (it is not so below 3 keV, because photo-absorption becomes increasingly important), and can thus serve as a proxy for the luminosity of the primary source. Recently, we have applied a similar technique to calculate the SS 433's reflected emission based on available data on the atomic and molecular gas distribution in the region of interest \citep{2016MNRAS.457.3963K}.

%\np
Although SS 433 is located $\sim 200$ pc out of the Galactic plane ($l,b=39.7\deg,-2.2\deg$ in Galactic coordinates) at the distance 4.5-5.5 kpc, \citep{2004ASPRv..12....1F,2004ApJ...616L.159B,2013ApJ...775...75M,2014A&A...562A.130P}, the direction of jets is nearly perpendicular to it with an intersection point at $ l\approx 39\deg$ \citep{2011MNRAS.414.2838G,2016MNRAS.457.3963K}.  

Since the atomic and molecular gas density distribution along the normal to the Galactic plane is characterized by an exponential profile  \citep{2009ARA&A..47...27K}, one might expect the corresponding reflection signal to have a peak around $ b=0\deg$, with its extent along Galactic longitude determined by the collimation angle of the illuminating emission \citep{2016MNRAS.457.3963K}.  

Because of that, we used a patch of the Galactic plane located around the intersection point ($38\deg<l<40\deg$, $|b|<0.25\deg$) and devoid of bright point sources to put a constraint on the reflected signal based on \textit{RXTE} \citep{2006A&A...452..169R} and \textit{ASCA} \citep{2001ApJS..134...77S} data. Due to the poor angular resolution of the former and the very limited observational coverage for the latter, the obtained constraint is mainly provided by reflection on the relatively smoothly distributed atomic gas, and the upper limit on SS 433' angle-integrated luminosity between 2 and 10 keV proves to be $\sim 2\times 10^{39}$ erg/s. Although this formally leaves open the possibility for SS 433 being a ULX, it is unlikely to be among the brightest, canonical sources of the class. 

%\np
On the other hand, the distribution of the Galactic molecular gas is essentially clumpy, with compact ($\lesssim10$ pc) and dense ($\gtrsim100$ cm$^{-3}$) structures (molecular clouds) dominating its mass budget \citep{2007ARA&A..45..565M}. In \cite{2016MNRAS.457.3963K}, we constructed a sample of potentially-illuminated molecular clouds (see Table 1, where positions and physical properties of these clouds are listed\footnote{There was a misidentification in the previous paper for one of the clouds, G041.04--00.26, due to the presence of entries with identical IDs in the catalogues of \cite{2009ApJ...699.1153R} and \cite{2010ApJ...723..492R}.}), based on the catalogue of clouds detected in the Boston University - Five College Radio Astronomy Observatory (BU-FCRAO) Galactic Ring Survey, which exploits the $^{13}$CO $J=1 \rightarrow 0$ emission line as a molecular gas tracer \citep{2009ApJ...699.1153R,2010ApJ...723..492R}. Thanks to the velocity information, one can estimate the distances to these clouds based on the rotational curve of the Galaxy, while the HI self-absorption (HISA) technique can help break the ``near-far degeneracy'' of this estimate \citep{2009ApJ...699.1153R}. However, the uncertainties in the resulting distance estimates are typically larger or comparable to the expected extent of the illumination region in the Galactic plane. Therefore, the constraints that can be obtained from the corresponding reflection signals are necessarily \textit{conditional} with respect to the incidence of a particular cloud into the illumination region.

%\np
From the observational point of view, the use of \textit{RXTE} data for studying the reflection on molecular clouds is inefficient since their angular size is much smaller than the angular resolution available with these data. As a result, the potential signal is significantly smeared out and contaminated by nearby (in projection on the sky) bright sources. The coverage of the region of interest by \textit{ASCA}, provided mainly by its Galactic Plane Survey {($|b|\lesssim 0.4\deg$)}, is quite limited, so that only a couple of the clouds in our sample appear to fall inside. Also, sensitive serendipitous observations by \textit{ASCA}, \textit{Chandra} and \textit{XMM-Newton} are available for some of the clouds, but in this case one should take into account the actual morphology of the predicted signal since its characteristic spatial extent is comparable with the field-of-view size of these observatories. 

%\np
In this paper, we first predict the morphology of the expected reflection emission from individual clouds under some simplifying assumptions regarding the relative disposition of a cloud and SS 433 and taking into account the actual morphology of the cloud, which is derived from the molecular emission line data filtered in the velocity space. Then, we describe how these predictions would change if the relative disposition of the cloud in the illumination region were different from the simple assumption. Finally, we compare the predictions with archival data of \textit{Chandra} and obtain constraints on the isotropic equivalent of the apparent face-on luminosity of the putative collimated emission of SS 433.

%####################################################### 
\begin{table*}
% origin : /home/ikh/Dropbox/ssrefl/paper/mc/mc15-topcat
\caption{The sample of molecular clouds that can potentially be illuminated by SS 433's collimated emission, constructed from the catalogues of \cite{2009ApJ...699.1153R} and \cite{2010ApJ...723..492R}. The clouds are described as elliptical regions in Galactic coordinates with centres at $(l_{MC},b_{MC})$ and half-axes $(\Delta l,\Delta b)$. Other columns are: line-of-sight distance ($d_{MC}$) with the corresponding $1\sigma$ uncertainty ($\delta d_{MC}$), characteristic radius ($r$), column density ($N_{H_2}$), total mass ($M_{H_2}$) and its uncertainty ($ \delta M_{H_2}$), and the minimal distance from SS 433 to a given cloud divided by the fiducial scale 200 pc. The clouds with \textit{Chandra} data available are marked in bold face.
\newline}
\begin{tabular}{clrrrrrlrcccc}
\hline\hline
  \multicolumn{1}{c}{} &
  \multicolumn{1}{c}{GRSMC} &
  \multicolumn{1}{c}{$l_{MC}$} &
  \multicolumn{1}{c}{$\Delta l$} &
  \multicolumn{1}{c}{$b_{MC}$} &
  \multicolumn{1}{c}{$\Delta b$} &
  \multicolumn{1}{c}{$d_{MC}$} &
  \multicolumn{1}{c}{$\delta d_{MC}$} &
  \multicolumn{1}{c}{r} &
  \multicolumn{1}{c}{N$_{H_2}$} &
  \multicolumn{1}{c}{$M_{H_{2}}$} &
  \multicolumn{1}{c}{$\delta M_{H_{2}}$} &
  \multicolumn{1}{c}{\underline{$R_{min}$}} \\
  \multicolumn{1}{c}{} &
  \multicolumn{1}{c}{name} &
  \multicolumn{1}{c}{deg} &
  \multicolumn{1}{c}{deg} &
  \multicolumn{1}{c}{deg} &
  \multicolumn{1}{c}{deg} &
  \multicolumn{1}{c}{kpc} &
  \multicolumn{1}{c}{kpc} &
  \multicolumn{1}{c}{pc} &
  \multicolumn{1}{c}{$10^{22}$cm$^{-2}$} &
  \multicolumn{1}{c}{$10^{4}M_{\odot}$} &
  \multicolumn{1}{c}{$10^{4}M_{\odot}$} &
  \multicolumn{1}{c}{200pc} \\  
\hline
  1  & G039.29--00.61 & 39.29 & 0.54 & -0.61 & 0.18 & 4.43 & 0.21 &  6.7 & 1.61 &  1.6 &  0.6 & 0.7\\
\textbf{2} &\textbf{G039.34--00.31} &\textbf{39.34} &\textbf{0.72} &\textbf{-0.31} &\textbf{0.24} &\textbf{4.55} &\textbf{0.18} &\textbf{7.7} &\textbf{1.91} &\textbf{2.5} &\textbf{1.0} &\textbf{0.8}\\
 \textbf{3} &\textbf{G041.04--00.26} &\textbf{41.04} &\textbf{0.44} &\textbf{-0.26} &\textbf{0.36} &\textbf{4.72} &\textbf{0.17} &\textbf{5.6} &\textbf{1.14} &\textbf{1.6} &\textbf{0.4} &\textbf{1.0}\\
  4  & G036.44+00.64 & 36.44 & 0.25 &  0.64 & 0.13 & 4.82 & 0.20 &  6.7 & 3.61 &  3.6 &  1.2 & 1.8\\
  5  & G036.39+00.84 & 36.39 & 0.40 &  0.84 & 0.19 & 4.80 & 0.15 &  5.0 & 2.17 &  1.2 &  0.4 & 1.9\\
  6  & G036.54+00.34 & 36.54 & 0.23 &  0.34 & 0.41 & 4.85 & 0.14 &  1.6 & 0.67 & 0.04 & 0.02 & 1.7\\
 \smallskip
\textbf{7} &\textbf{G039.34--00.26} &\textbf{39.34} &\textbf{0.58} &\textbf{-0.26} &\textbf{0.99} &\textbf{4.93} &\textbf{0.46} &\textbf{9.0} &\textbf{1.92} &\textbf{3.4} &\textbf{1.1} &\textbf{0.9}\\
  8   & G039.04--00.91 & 39.04 & 0.60 & -0.91 & 0.55 & 5.10 & 0.42 &  8.2 & 1.97 &  2.9 &  1.1 & 0.7\\
\textbf{9} &\textbf{G036.14+00.09} &\textbf{36.14} &\textbf{0.20} &\textbf{0.09} &\textbf{0.16} &\textbf{5.15} &\textbf{0.25} &\textbf{2.8} &\textbf{1.00} &\textbf{0.17} &\textbf{0.07} &\textbf{1.9}\\
  10 & G036.09+00.64 & 36.09 & 0.30 &  0.64 & 0.31 & 5.20 & 0.27 &  9.8 & 2.59 &  5.4 &  1.6 & 2.1\\
  11 & G037.74--00.46 & 37.74 & 0.88 & -0.46 & 0.43 & 5.25 & 0.33 &  8.8 & 1.04 &  1.8 &  0.7 & 1.2\\
  12 & G040.34--00.26 & 40.34 & 0.46 & -0.26 & 0.39 & 5.43 & 0.70 & 10.5 & 2.08 &  5.1 &  1.3 & 1.0\\
  13 & G041.24+00.39 & 41.24 & 0.63 &  0.39 & 0.61 & 5.53 & 0.41 &  3.6 & 0.98 &  0.3 &  0.1 & 1.5\\
  14 & G037.69--00.86 & 37.69 & 0.62 & -0.86 & 0.40 & 5.60 & 0.27 &  5.6 & 1.35 &  1.0 &  0.4 & 1.2\\
  15 & G036.89--00.41 & 36.89 & 0.42 & -0.41 & 0.37 & 5.70 & 0.32 & 16.0 & 2.18 & 12.0 &  4.4 & 1.7\\
\hline\end{tabular}

%  379 & G041.04-00.26 & 41.04 & -0.26 & 65.82 & 0.44 & 0.36 & 5.16 & 4.72 & 0.17 & 2.3977067604813644 & 5.6 & 16000.0 & 4000.0 & 327.0 & 6.17 & 1.37 & 159.2 & 1.41 & i & RD09 & 41.0993 & -0.34162582199999997 & 63.70856008184 & 1.7240366917 & 10 & 76760 & 0.5275355336095254 & 0.9877589812660985 & 3.0458159064882175 & 1.0848191556895261 & 3.3041594400622767 & 17485.5702102145\\

\label{t:mcs}
%}
\end{table*}
%#######################################################
%%%%%%%%%%%%%%%%%%%%%%%%%%%%%%%%%
\section{2. Predicted X-ray reflection signal}
\label{s:model}
%%%%%%%%%%%%%%%%%%%%%%%%%%%%%%%%%

%%%%%%%%%%%%%%%%%%%%%%%%%%%%%%%%%
\subsection{2.1. X-ray reflection}
\label{ss:reflection}
%%%%%%%%%%%%%%%%%%%%%%%%%%%%%%%%%
%

Interaction of X-ray emission with cold atomic or molecular gas has been thoroughly studied both from theoretical \citep{1996AstL...22..648S} and observational points of view \citep{2017MNRAS.465...45C,2017MNRAS.471.3293C}, and elaborated methods have been developed and applied in a broad variety of astrophysical contexts. In application to properties of the reflected emission, the main involved processes are photoabsorption by predominantly neutral metal atoms, Compton scattering on the electrons in hydrogen atoms and molecules, and fluorescence emission following photoabsorption \citep{1996AstL...22..648S}.

A number of useful simplifications can be applied to the problem in hand. First, the intensity of the incident X-ray radiation is too low for the thermal, ionization or chemical state of the illuminated gas to be affected, for any reasonable value of SS 433's apparent luminosity \citep{2016MNRAS.457.3963K}. Therefore, the resulting reflected emission should scale linearly with intensity of the illuminating radiation. 

Second, for structures characterized by hydrogen column density $N_{H}\lesssim 10^{23}$ cm$^{-2}$ the impact of photoabsorption remains small for photons with energies above $\gtrsim 3$ keV, while the optical depth with respect to Compton scattering does not exceed 0.1 either. As a result, calculations in the optically thin limit should be sufficiently accurate, taking into account much larger uncertainties inherent to other aspects of the problem (e.g. in relative disposition of the primary source and the reflector or variations in the gas metallicity). Consequently, the reflection signal simply scales linearly with the mass of the illuminated gas. 

Finally, the geometry of the problem (namely, the fact that the assumed axis of the collimated emission is nearly perpendicular to the plane of the Galactic disc, where the molecular clouds of interest are located) implies that $90\deg$-scattering is likely taking place in the majority of cases\footnote{Interestingly, this also implies a high polarisation degree of the reflected emission, which might be used as a proof against other scenarios for the emission origin, e.g. illumination by cosmic rays or an X-ray source inside the cloud \citep{2002MNRAS.330..817C,2017MNRAS.468..165C}.}. Because of that, the efficiency of reflection can be described by a single energy-independent albedo factor determined mainly by the scattering cross-section on electrons bound in hydrogen atoms and molecules (including contributions of Rayleigh, Raman and Compton scattering), which turns out to be approximately equal to the cross-section for Thomson scattering by free electrons ($ \sigma_T=6.65\times10^{-25}$ cm$^2$, as far as relativistic corrections can be neglected) \citep{1996AstL...22..648S}. 

Similarly, one can take the contribution of helium atoms and fluorescent lines into account, resulting in the effective reflection cross-section (per hydrogen) equal to $ \sigma_{eff}=1.7\sigma_T$ for an approximately solar abundance of the elements, if the 4-8 keV energy band is considered \citep{2017MNRAS.471.3293C}. This band, being practically unaffected by interstellar absorption and containing the brightest fluorescent line, i.e. the iron 6.4 keV line, has proved to be optimal for X-ray reflection studies based on data of the \textit{Chandra} and \xmm~ observatories \citep{2017MNRAS.465...45C} and will also be used actively in the present study.

With all this in mind, the \textit{apparent} 4-8 keV luminosity of a reflector of $H_2$ mass $ M_{H_2}=10^4M_{4}\, M_{\odot}$ located at the distance $ R=200\,R_{200}$ pc from the illuminating source is given by
\begin{equation}
L_{sc,4-8}=\frac{\sigma_{eff}}{4\pi R^2}\frac{M_{H_2}}{\mu_p\,m_p}L_{4-8}\approx2\times 10^{33} \frac{L_{39}}{R_{200}^2}~{M_4}~{\rm erg/s},
\label{eq:lumsc}
\end{equation}
where $L_{4-8}=10^{39}L_{39}$ erg/s is the isotropic equivalent of the apparent luminosity, $ \mu_{p}=1.4$ is the mean molecular weight per hydrogen. 

The resulting 4-8 keV flux of the reflected emission (at distance $d_{MC}=5d_{5}$ kpc) can be evaluated as 
\begin{equation}
\label{eq:fh2sc}
f_{H_2,4-8}=6.7\times 10^{-13} \frac{L_{39}}{d_5^2 R_{200}^2}{M_4}~{\rm erg/s/cm^2}.
\end{equation} 

The slight difference of Eqs. (1) and (2) above from to Eqs. (13) and (21) in \cite{2016MNRAS.457.3963K} is due to the narrow energy range considered here, for which the contribution of the iron fluorescent line results in an increased effective scattering cross-section (see \cite{2017MNRAS.471.3293C}). 

%%%%%%%%%%%%%%%%%%%%%%%%%%%%%%%%%
\subsection{2.2. Molecular gas distribution}
\label{ss:molecular}
%%%%%%%%%%%%%%%%%%%%%%%%%%%%%%%%%

%%%
\begin{figure}
\includegraphics[width=1.\columnwidth, bb=50 180 540 660]{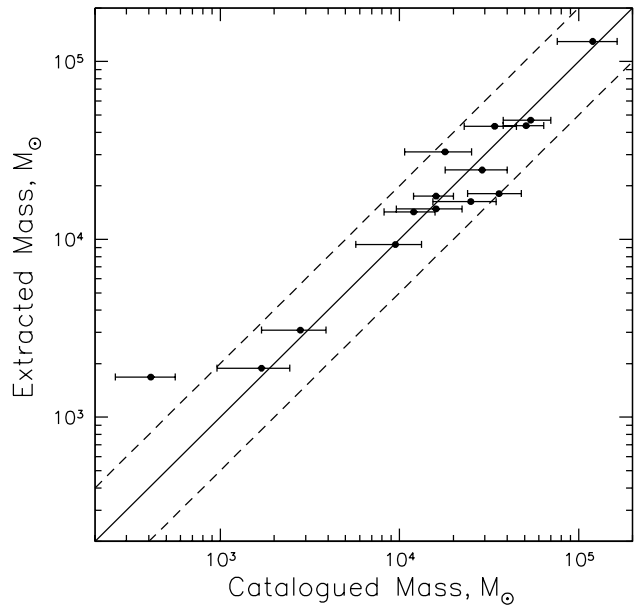}
\caption{Consistency of the molecular cloud masses listed in the catalogue of \cite{2010ApJ...723..492R} (also presented in Table 1 with the corresponding uncertainties) and the masses of the molecular gas extracted from the original data cubes according to the procedure described in Section 2.2. The dashed lines mark a factor of two difference between them. We see that for all the clouds except for the least massive one, the masses are consistent within this factor.}
\end{figure}
%%%

%%%%%%%%%%%%%%%%%%%%%%%%%%%%%%%%%%%%%%%%%%%%%%%%%%%%%%%%%%%%%
\begin{figure*}
\includegraphics[width=1.\textwidth, bb= 50 335 550 530]{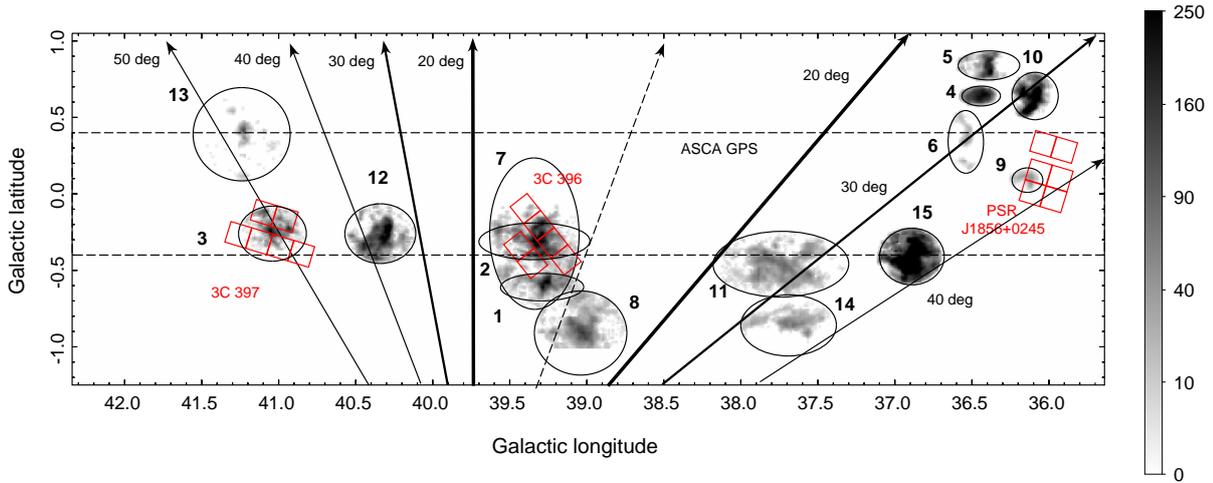}
\caption{
Projected mass of the molecular gas (per 44''$\times$ 44'' pixel, $M_{\odot}$ units) associated with the clouds that can potentially be illuminated by SS 433's collimated X-ray emission (listed in Table \ref{t:mcs}). Boundaries of the clouds are depicted as black ellipses. Footprints of \textit{Chandra} observations analysed in this paper are overlaid in red. Also shown is the boundary of the coverage region of the \textit{ASCA} Galactic Plane Survey ({$|b|\lesssim 0.4\deg$}, \citep{2001ApJS..134...77S}) and projections of cones (solid lines) with the axis along the jets' precession axis (thick dashed line) and various opening angles (as labelled next to each line).
} 
\label{f:mass}
\end{figure*}
%%%%%%%%%%%%%%%%%%%%%%%%%%%%%%%%%%%%%%%%%%%%%%%%%%%%%%%%%%%%%

In contrast to the atomic gas, molecular gas is concentrated in dense clouds that are most likely smaller ($\lesssim 15$ pc) than the extent of the potential illumination region ($\lesssim 100$ pc). Because of that, the reflected signal expected in this case depends crucially on the incidence of such clouds inside this region, determined by its geometry and relative disposition of the source and the clouds (see  \cite{2016MNRAS.457.3963K} for an illustration and thorough discussion). Unfortunately, the relative disposition cannot be firmly established due to significant uncertainties in the line-of-sight distances both to SS 433 and the molecular clouds, which turn out to be comparable with the expected extent of the illumination region in the Galactic plane. Current estimates of the distance to SS 433 vary from 4.5 to 5.5 kpc \citep{2004ApJ...616L.159B,2013ApJ...775...75M,2014A&A...562A.130P}.

Besides that, these clouds are also characterized by a wide diversity of (irregular) morphological shapes and internal structures. One can see that typical angular sizes of the clouds ($\sim 10$ arcmin) are comparable to the field-of-view (FOV) size of \textit{Chandra} and \textit{XMM-Newton} observatories, so their actual morphology has to be taken into account in order to correctly predict the distribution of the extended reflection signal across the observational domain. 

For this purpose, one needs to take advantage of the original data of the BU-FCRAO Galactic Ring Survey\footnote{https://www.bu.edu/galacticring/new\_data.html} for the region of interest, i.e. for Galactic longitude spanning from $ l=36^{\deg} $ to $ l=42^{\deg}$, Galactic latitude from $ b=-1^{\deg}$ to $ b=1^{\deg}$, and line-of-sight velocities $ V_{lsr} $ from 60 to 80 km/s (see \citealt{2016MNRAS.457.3963K} for the details). These data are presented in the form of background-subtracted antenna temperatures $T_a$ for the $^{13}$CO line emission measured on a position-position-velocity (PPV) coordinate grid \citep{2009ApJS..182..131R}. We re-binned the original data cubes to twice worse sampling in all dimensions, which resulted in pixel sizes $ \delta l=\delta b=44$ arcsec along Galactic longitude and latitude, and $\delta V_{lsr}=$0.425 km/s in $ V_{lsr}$ dimension. Since we are interested in the total reflected emission from the clouds with characteristic sizes of $ \sim 10 $ arcmin and velocity dispersion of $\Delta V_{lsr}\sim $ few km/s (see Table 1 and \citealt{2016MNRAS.457.3963K} for $\Delta V_{lsr}$), such a sampling is fully appropriate for the purposes of the current study.

We filtered the original data cube to leave only voxels that fall inside the PPV regions of the molecular clouds in our sample. Namely, these regions are 3D ellipsoids in PPV space, centred on $(l_{MC},b_{MC})$ and  $V_{LSR,MC}$ (see Table 1). The ellipsoids' major axes are $\Delta l$ and $\Delta b$ along the Galactic coordinates and $\Delta V_{LSR}$ along the line-of-sight velocity coordinate. On top of that, we only left voxels with the signal-to-noise ratio higher than 3 to minimize the influence of the background-subtraction on the inferred masses.

After correction for the main beam efficiency, $T_{mb}=T_a/0.48$ (e.g. \cite{2010ApJ...723..492R}), one can compute the integrated line intensities for each data voxel $I_{mb}=T_{mb}\delta V $ (in K km/s units). The corresponding column density of the molecular gas for a gas voxel is calculated in the standard way \citep{2009ApJS..182..131R}:
\beq
N_{H_2}=4.92\times 10^{20} \mathrm{cm^{-2}} I_{mb}.
\label{eq:nh2vox}
\eeq
The corresponding molecular mass is given by
\beq
m_{H_2}=2 m_p N_{H_2}~\delta x ~\delta y,
\label{eq:mvox}
\eeq
where $ \delta x\approx d_{MC} \delta l$ and $ \delta y\approx d_{MC}\delta b  $, with  $ d_{MC}$ being the distance to the molecular cloud. For $ d_{MC}\approx 5$ kpc, one has $ \delta x=\delta y\approx 1.07$ pc, so combining Eq. \ref{eq:mvox} and Eq. \ref{eq:nh2vox} yields
\beq
m_{H_2}=7.93 M_{\odot}\times T_{a} \left(\frac{d_{MC}}{5 \mathrm{kpc}}\right)^2.
\label{eq:mvoxta}
\eeq

After correction for the actual mean excitation temperature measured for each cloud, we checked that the resulting masses for the individual clouds are consistent with those cited in the catalogue of \cite{2010ApJ...723..492R} (see Figure 1). The only exception is the least massive cloud G036.54+00.34, which, however, is of no importance for the current study since it has not been covered by any \textit{Chandra} or \textit{XMM-Newton} observation.

%%%%%%%%%%%%%%%%%%%%%%%%%%%%%%%%%
\subsection{2.3. Simulation}
\label{ss:simulation}
%%%%%%%%%%%%%%%%%%%%%%%%%%%%%%%%%
 
As we noted previously, the key aspect for the illumination of molecular clouds is their disposition with respect to the primary source of the collimated emission. We modelled the spatial distribution of the molecular gas extracted from the original data cubes (as described in the previous Section) by calculating the line-of-sight distance to each voxel based on its line-of-sight velocity via the standard kinematic distance formulae, given by Eqs. (7) and (8) in \cite{2016MNRAS.457.3963K}. As a result, we obtained a set of voxels with 3D positions and enclosed molecular mass. The expected X-ray reflection signal was then obtained by combining signals from individual voxels calculated according to Eqs. (1) and (2).

Since the illuminating radiation of interest is collimated, one needs to model the incidence of a particular cloud in the illumination region. The overall geometry of this region with allowance for possible precession of the radiation cone axis has been fully described in \cite{2016MNRAS.457.3963K}. The 3D direction of the precession axis can be fully determined thanks to the spatially resolved precession pattern of radio emission at arcsec scales \citep{2004ApJ...616L.159B,2008ApJ...682.1141M} and the detection of Doppler-shifted emission lines from the arcsec-scale X-ray emission of SS 433  \citep{2002Sci...297.1673M,2017AstL...43..388K}. As a result, the location and geometry of the illumination region is fully determined by two parameters: the half-opening angle of the collimated emission cone, $\Theta_r$, and the position of the illuminating source along the line-of sight, i.e. the distance to SS 433 $d_{SS433}$.

Here, we focus mainly on the situation where the collimation angle is close to the maximal one allowed by non-visibility of the collimated emission by an observer on Earth taking into account precession of the cone axis with amplitude $\Theta_p=21\deg$ and nutation with amplitude $\Theta_n\sim 5\deg$, i.e. $\Theta_{r,max}\sim i-\Theta_{p}-\Theta_{n}\sim 50\deg$ \citep{2004ASPRv..12....1F} . Clearly, in this case the dependence of the predicted signal on distance to SS 433 is minimized. In addition, we performed similar calculations for smaller collimation angles, in application to the particular clouds for which X-ray data are available.

In order to model the dependence on the position of the illuminating source, we performed calculations for a grid of distances ranging from 4.5 to 5.5 kpc, assuming that the reconstructed 3D distribution of the molecular gas is precise (this is actually not true, see the distance uncertainties listed in Table 1 and Section 3.2 in \cite{2016MNRAS.457.3963K}). Thus, the adopted range of distances is actually meant to represent the combined uncertainty in the relative disposition of SS 433 and the clouds.

In Figure 3, we illustrate the predicted surface brightness of the reflected emission from the source with apparent 4-8 keV luminosity $L_{4-8}=10^{39}$ erg/s and collimation angle 50$\deg$ for three values of its distance: 4.5, 5.0 and 5.5 kpc. Naturally, the clouds located closer to the primary source turn out to be brightest in the reflected emission. Note that we do not take into account the 'duty-cycle' of illumination, i.e. the fraction of time a given direction spends inside the precessing illumination cone \citep{2016MNRAS.457.3963K}, since practically no constraint can be imposed on the corresponding correction factor. Hence, for a given apparent luminosity of the primary source, the actual reflection signal may be somewhat weaker than predicted here.

The predicted picture is fully consistent with the simple estimates presented in Table 2 of \cite{2016MNRAS.457.3963K}, which were based on averaged characteristics of the clouds.The typical expected surface brightness of the reflected emission for the fiducial incident luminosity quoted above is $\sim $few$\times 10^{-11}$ erg/s/cm$^2$/deg$^2$, reaching  $\sim 10^{-10}$ erg/s/cm$^2$/deg$^2$ in the brightest regions. It is important to notice that the X-ray background in the region of interest is measured to be $\sim2\times 10^{-11}$ erg/s/cm$^{2}$/deg$^2$, with approximately equal contributions from the cosmic (i.e. extragalactic) X-ray background (CXB) and the Galactic X-ray ridge emission (GRXE) \citep{2001ApJS..134...77S}.

Therefore, on the one hand, the predicted signal should be well detectable for an apparent illuminating luminosity of $10^{39}$  erg/s. On the other hand, it is clear that in the case of non-detection, the upper limit on the sought signal will be primarily determined by the accuracy of background subtraction. Unfortunately, the actual background count rate turns out to be even higher due to the imperfect subtraction of the particle background, and it is the precision of determination of this component that sets the actual sensitivity for faint extended sources such as the X-ray reflection signals we are looking for. 

%%%%%%%%%%%%%%%%%%%%%%%%%%%%%%%%%%%%%%%%%%%%%%%%%%%%%%%%%%%%
\begin{figure*}
\includegraphics[width=1.\textwidth, bb= 50 340 530 530]{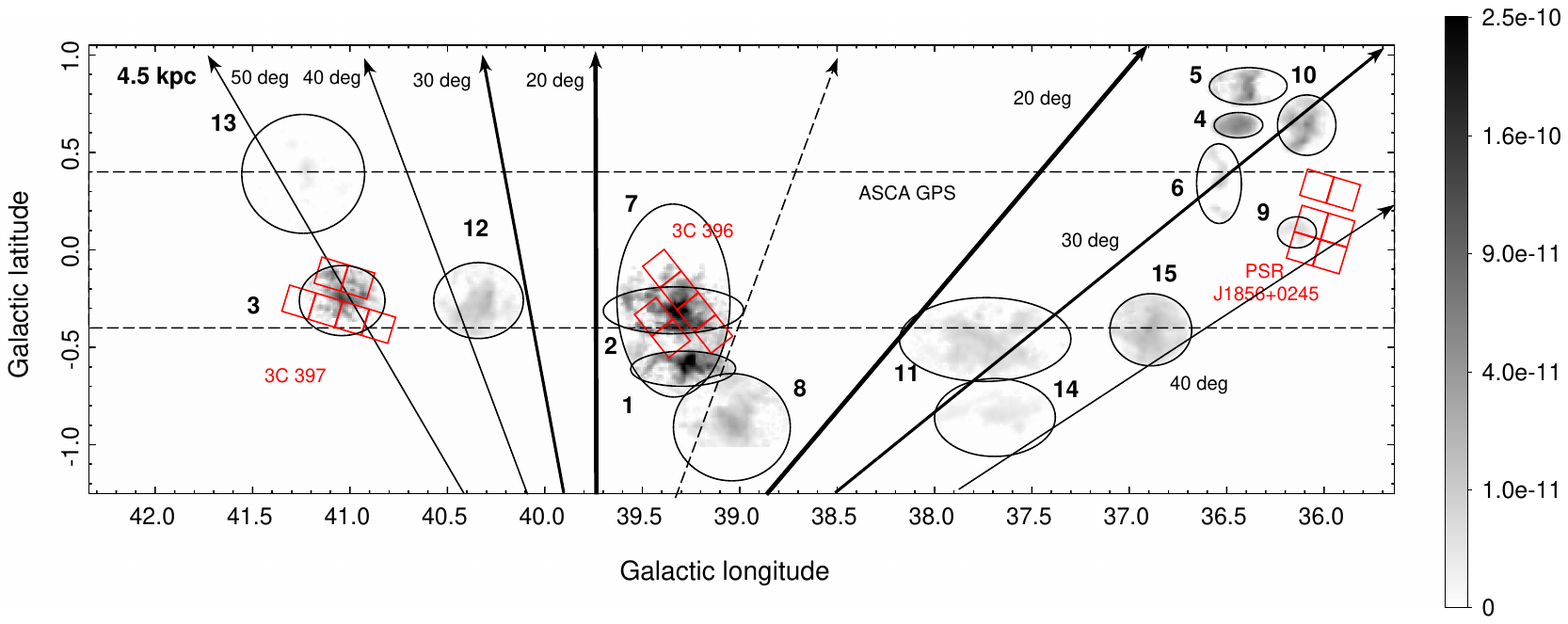}
\includegraphics[width=1.\textwidth, bb= 50 340 530 530]{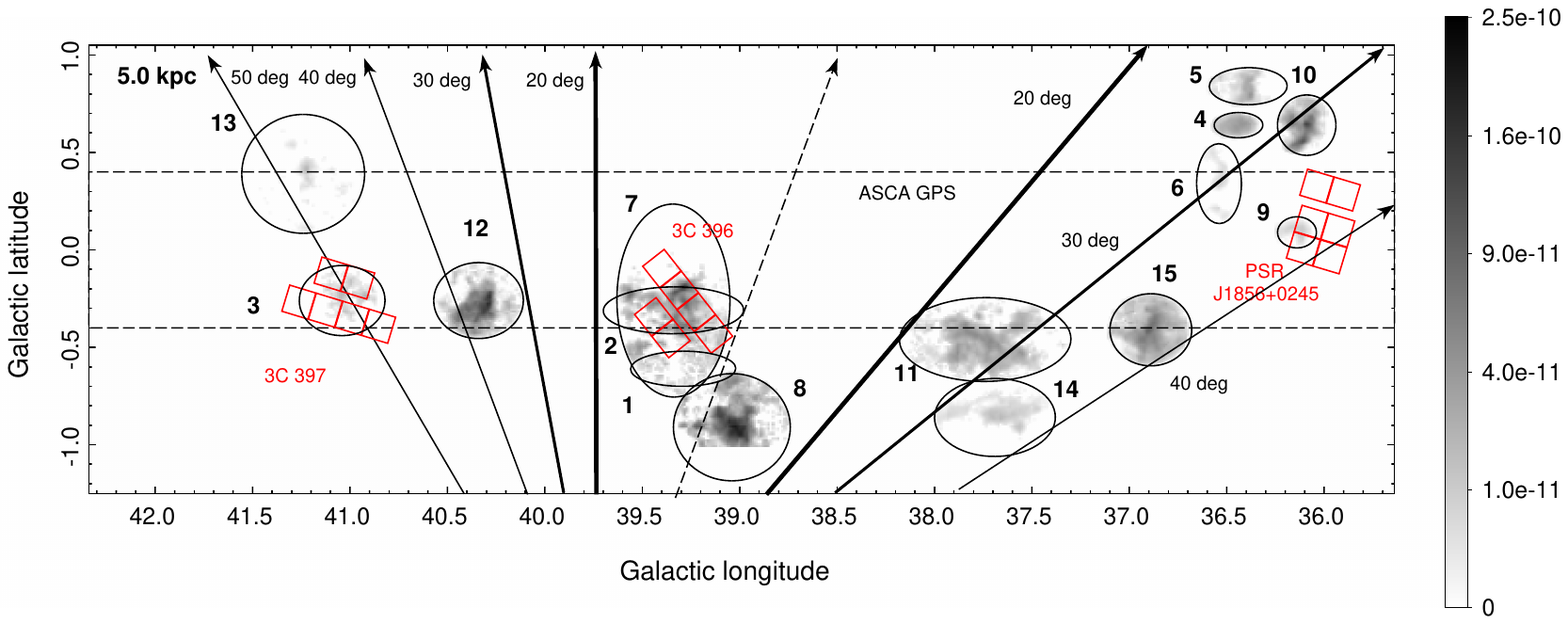}
\includegraphics[width=1.\textwidth, bb= 50 340 530 530]{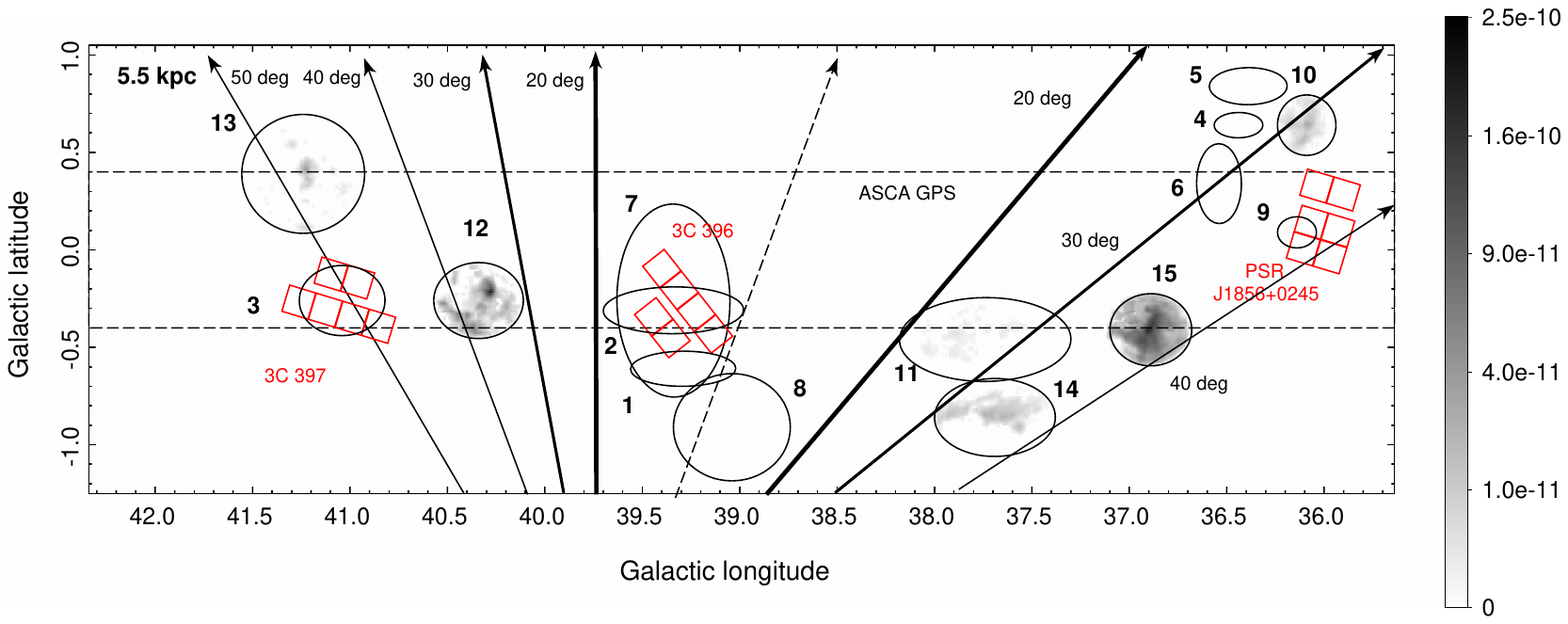}
\caption{Predicted surface brightness (erg/s/cm$^2$/deg$^2$) of the reflected emission in the 4-8 keV energy range for the corresponding luminosity of SS 433 equal to $10^{39}$ erg/s and a collimation angle of 50 degrees. The distance to SS 433 is assumed to be 4.5, 5 and 5.5 kpc in the top, middle and bottom panel, respectively. The overlaid regions are the same as in Figure 2.} 
\label{f:sb}
\end{figure*}
%%%%%%%%%%%%%%%%%%%%%%%%%%%%%%%%%%%%%%%%%%%%%%%%%%%%%%%%%%%%%
  
%%%%%%%%%%%%%%%%%%%%%%%%%%%%%%%%%
\section{3. X-ray data}
\label{s:data}
%%%%%%%%%%%%%%%%%%%%%%%%%%%%%%%%%
%
\subsection{Data set}

We searched the \textit{Chandra} Data Archive for all publicly available observations that cover regions where one might find the reflected emission predicted in Section 2. Three observations were found: ObsIDs 1042, 1988, 12557. The primary targets of these observations were supernovae remnants 3C 397 and 3C 396, and the pulsar PSR J1856+0245, respectively. 

The corresponding molecular clouds are G039.34-00.31 and  G039.34-00.26 (clouds 2 and 7 in Table 1) for ObsID 1988, G041.04-00.26 (cloud 3) for ObsID 1042, and G036.14+00.09 (cloud 9) for ObsID 12557. All these clouds are marked with bold face in Table 1. The actual distribution of the predicted reflected emission across the telescope's field of view is demonstrated in Figure 3 for three different values of distance to SS 433 (4.5, 5, and 5.5 kpc). The significant variations visible across the covered regions clearly demonstrate the importance of accurate modelling of the predicted emission morphology.

\subsection{Data preparation and analysis}

The initial data processing was performed using the latest calibration data and following the standard procedure described in \cite{2005ApJ...628..655V}. We performed corrections for the exposure, vignetting effect and the particle background according to the procedures used in \cite[][]{2012MNRAS.421.1123C,2017MNRAS.471.3293C}. 

This allowed us to construct flattened maps of the measured count rate per pixel in the spectral band of interest, i.e. 4-8 keV. We ensured that the maps were not  contaminated by bright extended sources, e.g. supernovae remnants 3C 396 (the primary target of ObsID 1988) and 3C 397 (the primary target for ObsID 1042). Namely, the aim-point chips were not used. Additionally, 12-arcsec radius circles centred on the positions of known point sources were excluded.

We produced the broad-band response functions (RMF and ARF) and used these to calculate a conversion factor between the count rate and X-ray flux for a  fiducial spectral model, namely, an absorbed power law with $N_{H}=10^{22}$ cm$^{-2}$ and slope $\Gamma=2$. Using this factor, the count rate maps were  converted into the corresponding X-ray surface brightness maps (see the upper panels in Figures 4, 6 and 7). Visual inspection of these maps shows that no strong signal morphologically resembling the predicted X-ray reflection signal (see middle panels in Figures 4 and 6) is present for any of the observed regions.

It should be mentioned that the \textit{Chandra} background count rate in the 4-8 keV band is dominated by particle background. Therefore, if no strong additional signal is present on the images, their appearance is strongly influenced by fluctuations arising due to the imperfection of background subtraction. In surface brightness units, these fluctuations have an amplitude of $\sim 2\times10^{-10}$ erg/s/cm$^2$/deg$^2$ (for the original data pixelization smoothed with a 3'' Gaussain kernel), i.e. a factor of 10 higher than the expected contribution of the X-ray background, although this amplitude strongly diminishes after integration over regions of substantial extent ($\sim$ a few arcmin).

Having defined the source and background regions so that the signal-to-noise ratio expected based on our X-ray reflection modelling (as described in Section 2) was maximized, we extracted the source and background surface brightness spectra. In order to further improve subtraction of the particle background component, we consider the difference between the surface brightness spectra extracted from the source and background regions.

Clearly, in the case of a virtually identical X-ray background for the source and background regions and a very weak additional signal (i.e. the X-ray reflection we are searching for), the resulting difference spectrum should be close to Gaussian with a zero mean and fluctuations determined by the sum of the X-ray and particle background subtraction uncertainties. In such a case, it is meaningful to consider only an upper limit on the excess surface brightness, which is essentially set by these uncertainties. Once again, we fix the shape of the spectral model (namely, absorbed power law with $N_{H}=10^{22}$ cm$^{-2}$ and slope $\Gamma=2$) for this excess emission, and set an upper limit on its normalization, and consequently the 4-8 keV flux.
 
% 
%%%%%%%%%%%%%%%%%%%%%%%%%%%%%%%%%
\section{4. Results}
\label{s:results}
%%%%%%%%%%%%%%%%%%%%%%%%%%%%%%%%%

%#######################################################
\subsection{G039.34-00.31 \& G039.34-00.26}
%#######################################################

\textit{Chandra} observation ObsId 1988 has been taken in ACIS-S configuration with a total exposure time of  98 ks, and its primary target was supernova remnant 3C 396 \citep{2003ApJ...592L..45O}. In order to avoid contamination by the extended emission of this supernova remnant, we fully excluded the aim-point chip (S3) from our consideration. The resulting map of the 4-8 keV surface brightness is shown in Figure 4 (upper panel). No significant excess over the background level is visible. 

As mentioned above, in order to obtain upper limits on the reflected emission,  it is crucial to select appropriate regions for extraction of the source and background signals. Here, there are two clouds, G039.34-00.31 and G039.34-00.26, falling inside the field of view (see the middle panel in Figure 4 for the predicted morphology of the reflected emission for the illuminating source at $d_{SS433}=4.8$ kpc with an apparent luminosity $L_{X}=10^{39}$ erg/s and the maximal collimation angle). Due to this, we select two source regions, s1 and s2 (see Figure 4), within which the predicted reflected emission is contributed by the two clouds in different proportions, depending on the illuminating source's disposition and the collimation angle (see the bottom panel in Figure 4). As can be expected from the line-of-sight distances of G039.34-00.31 (4.5$\pm$0.2 kpc) and G039.34-00.26 (4.9$\pm$0.5 kpc), the predicted emission has a broad peak for the illuminating source's distances around $d_{SS433} \approx 4.8$ kpc for region \textit{s1} and $d_{SS433} \approx 4.65$ kpc for region \textit{s2}.

Besides that, the potential reflection signal is very extended, covering a significant portion of the whole aperture, so it is impossible to select background regions fully devoid of it. We nevertheless constructed three regions for which it is at least an order of magnitude weaker than for the source regions (see the blue curves in the bottom panel in Figure 4). 

Also, for apparent luminosities below $L_{X}=10^{39}$ erg/s, the reflection signal predicted inside the background regions is at least few times less than the expected contribution of the X-ray background. As a result, possible over-estimation of the background level due to this contamination (and hence possible background over-subtraction from the source regions) should be much less important than the actual uncertainty in background determination.
 
In order to estimate possible background variations from one of our regions to another (such as might be expected especially for the GRXE part of XRB, given the  $\approx 0.5\deg$ extent of the field of view along the Galactic latitude), we subtracted, for each source region, background spectra determined in three separate regions and calculated the surface brightness of the residual emission. The resulting values for both source regions turn out to be consistent with each other ($\sim10^{-11}$ erg/s/cm$^2$/deg$^2$) within the corresponding uncertainties ($\sim$few $\times10^{-12}$ erg/s/cm$^2$/deg$^2$) - see the left panel in Figure 5. We therefore combined these three values for each source region, which resulted in the measured residual surface brightness $(7\pm3)\times10^{-12}$ erg/s/cm$^2$/deg$^2$ for region 1 and $(11\pm4)\times10^{-12}$ erg/s/cm$^2$/deg$^2$ for region 2. The corresponding 1$\sigma$ upper limits on the reflected emission are hence $1.0\times10^{-11}$ erg/s/cm$^2$/deg$^2$ and $1.5\times1.0^{-11}$ erg/s/cm$^2$/deg$^2$, respectively.

Now we can compare these upper limits to the predicted surface brightness of the reflected emission from the same regions. For the illuminating source's apparent luminosity $10^{39}$ erg/s and a collimation angle of 50$\deg$, it is predicted to be  $\approx10^{-10}$ erg/s/cm$^2$/deg$^2$ over a broad range of distances to SS 433, from 4.6 to 5 kpc (see left panel in Figure 5). Hence, the obtained upper limits translate into an upper limit on the luminosity of $2\times 10^{38}$ erg/s for both source regions and all SS 433 distances less than 5 kpc. 

Taken at face value, this result would effectively exclude association of SS 433 with ULXs for half of its currently allowed range of distances, i.e. for $d_{SS 433}$ from 4.5 to 5 kpc. Such broadness of the derived constraint stems from the large assumed collimation angle. In order to illustrate how the situation changes for smaller collimation angles, we calculated the X-ray reflection also for $\Theta_r=21\deg$ (i.e. equal to the jets' precession amplitude $\Theta_p=21\deg$) and for $\Theta_r=10\deg$ (see the right panel in Figure 5). The resulting upper limits on the apparent luminosity do become weaker, but nevertheless $L_{X}<10^{39}$ erg/s for all $d_{SS433}$ below 5 kpc regardless of the collimation angle. 

The reason is that these two clouds lie inside the sky projection of the SS 433's precession cone (see Figure 2) and are quite extended (covering together $\sim 40$ pc along the line of sight, see Table 1), so even for relatively small collimation angles, $\Theta_r\sim10\deg$ there is a significant chance of some part of their mass falling inside the precessing illumination cone (see \cite{2016MNRAS.457.3963K} for a detailed discussion). Of course, for small collimation angles,  $\Theta_r\ll \Theta_{p}$, one needs to apply the correction for `duty-cycle' of the illumination \citep{2016MNRAS.457.3963K}, but this effect is smaller than the uncertainty caused by the lack of precise distance estimates for the clouds. 

%%%%%%%%%%%%%%%%%%%%%%%%%%%%%%%%%%%%%%%%%%%%%%%%%%%%%%%%%%%%%
\begin{figure*}
\center
%origin: /home/ikh/Dropbox/ssrefl/paper/figs/3d/angles.svg
\includegraphics[width=1.0\columnwidth,bb=50 200 600 630]{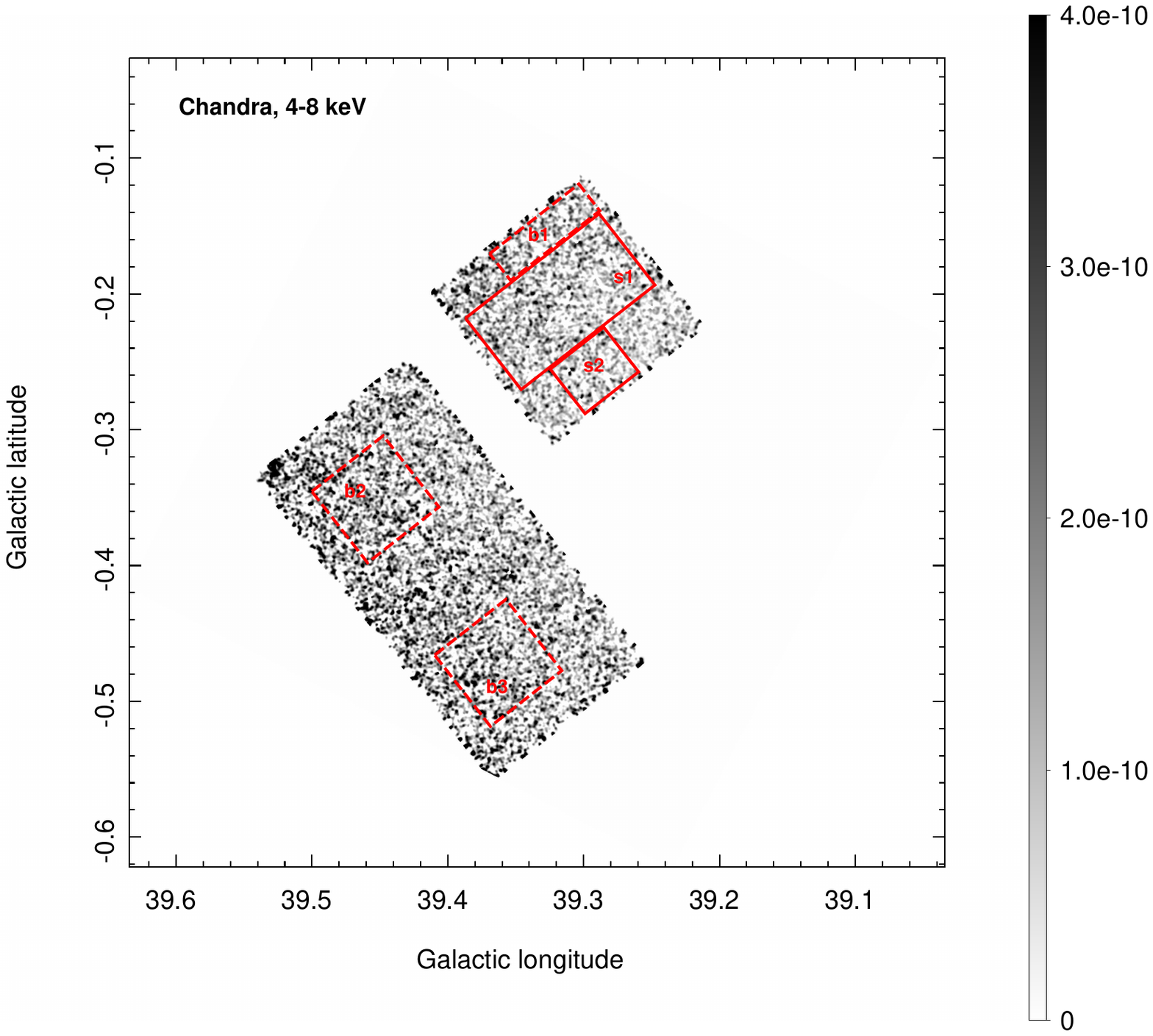}\\
\includegraphics[width=1.0\columnwidth,bb=50 200 600 630]{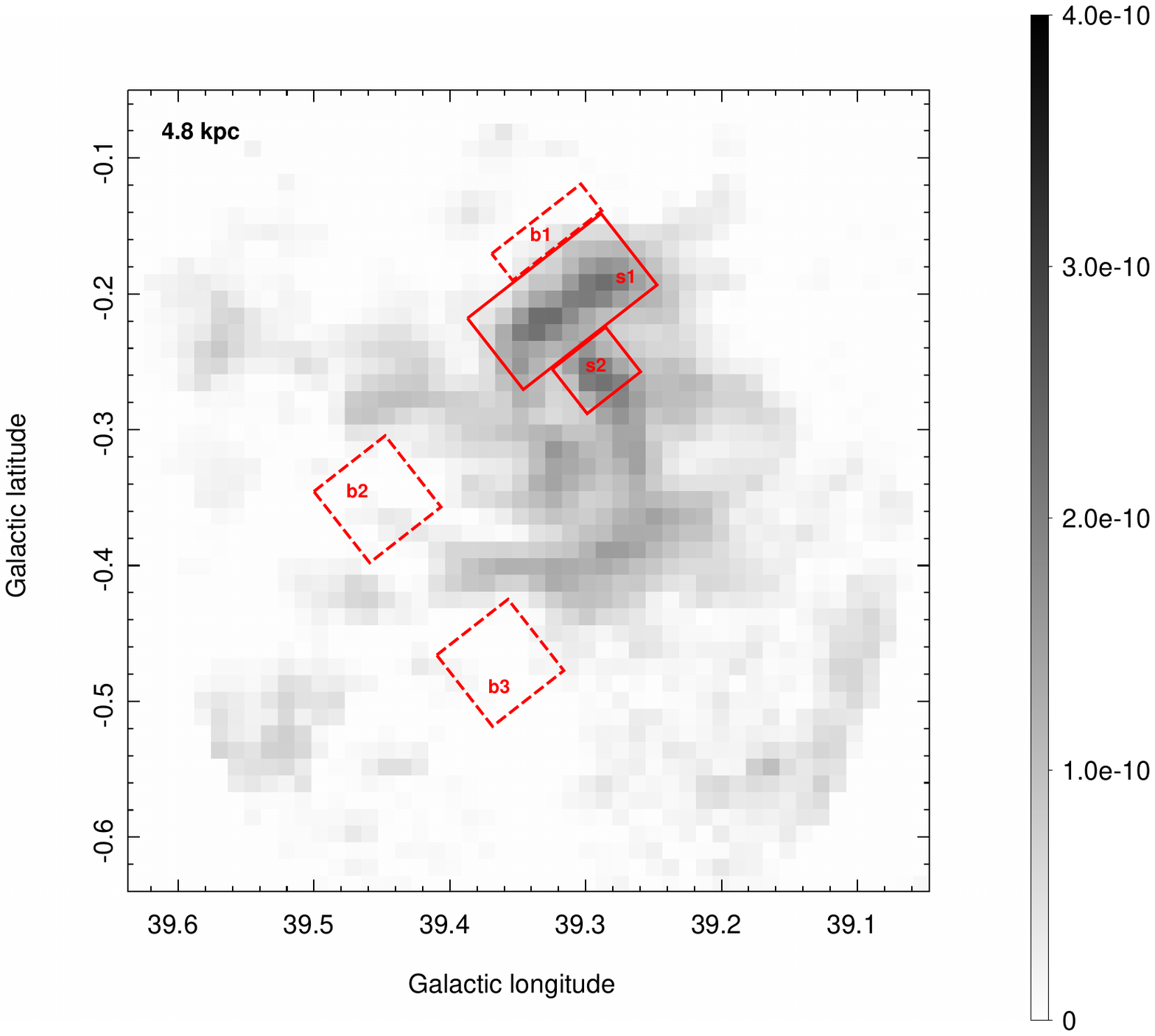}\\
\includegraphics[width=0.75\columnwidth,bb=100 200 600 680]{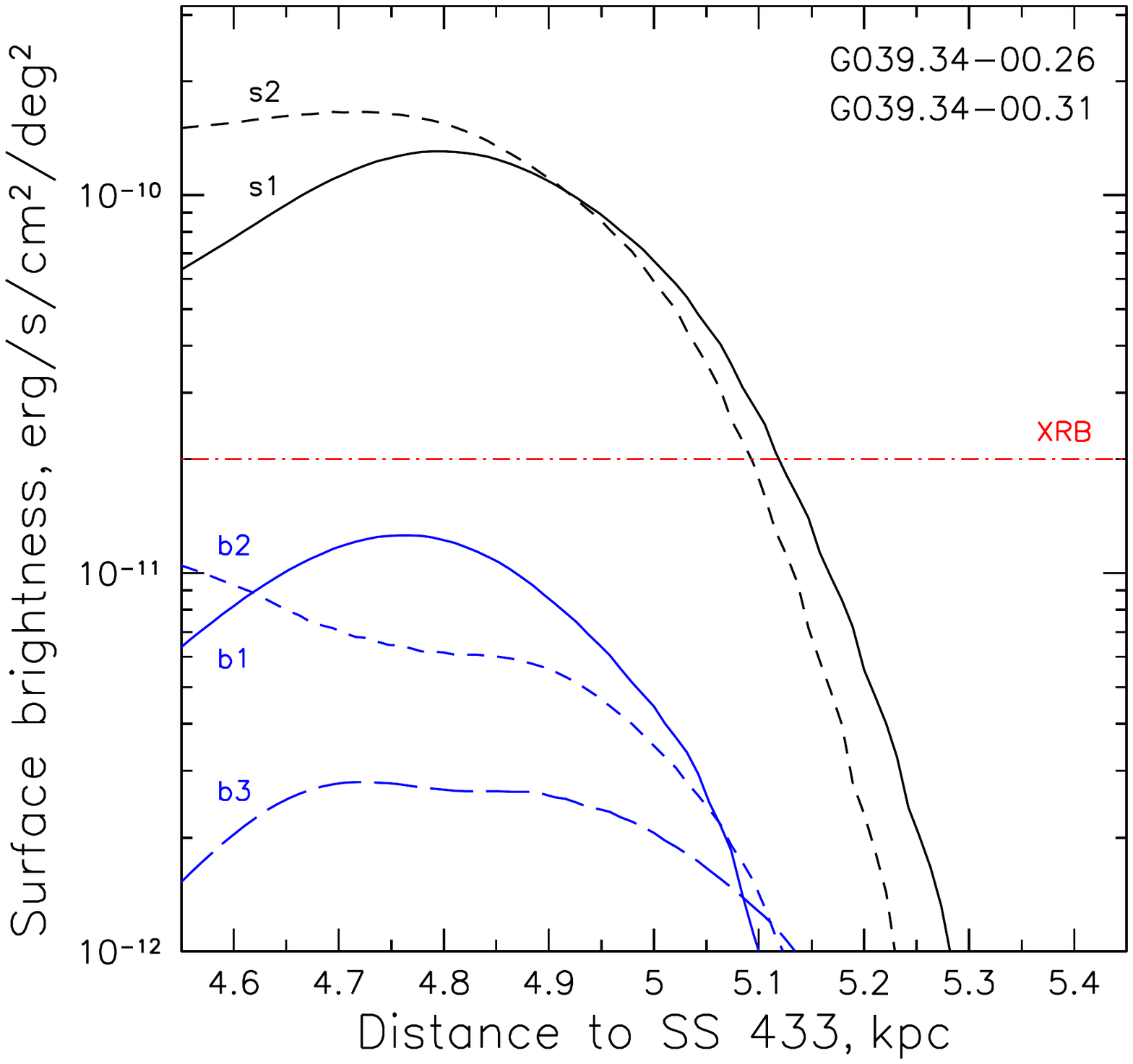}
\caption{The region where X-ray reflection on the clouds G039.34-00.31 and G039.34-00.26 is expected and which is covered by a \textit{Chandra} observation. \textit{Top panel.} The map of 4-8 keV surface brightness (erg/s/cm$^2$/deg$^2$) derived from \textit{Chandra} data after exclusion of point sources. The solid rectangular regions (labelled s1 and s2) were used as signal extraction regions, while the dashed regions (labelled b1, b2, and b3) were used as regions for background determination. \textit{Middle panel.} Predicted 4-8 keV surface brightness (erg/s/cm$^2$/deg$^2$) of the reflected emission for the primary source at 4.8 kpc with an apparent 4-8 keV luminosity of $10^{39}$ erg/s and a collimation angle of 50$\deg$. \textit{Bottom panel.} Dependence of the predicted surface brightness in these regions on the assumed distance to the illuminating source: black curves -- for the source regions, blue curves -- for the background regions (as indicated next to each curve). The red dash-dotted line shows the expected level of the X-ray background (CXB+GRXE).} 
\label{f:3C396}
\end{figure*}
%%%
\begin{figure*}
%origin: /home/ikh/Dropbox/ssrefl/paper/figs/3d/angles.svg
\includegraphics[width=\columnwidth,bb=50 150 600 700]{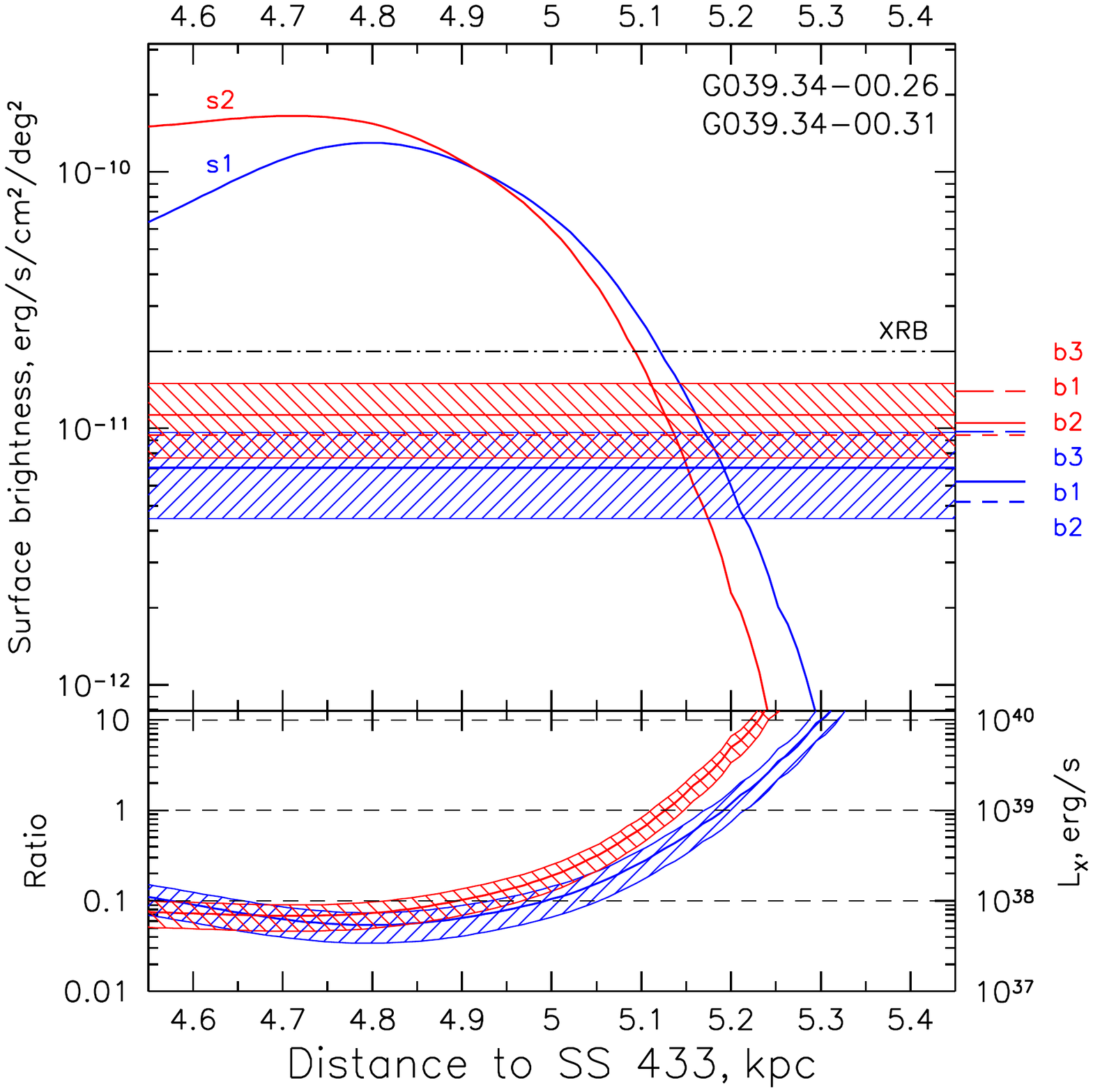}
\includegraphics[width=\columnwidth,bb=50 150 600 700]{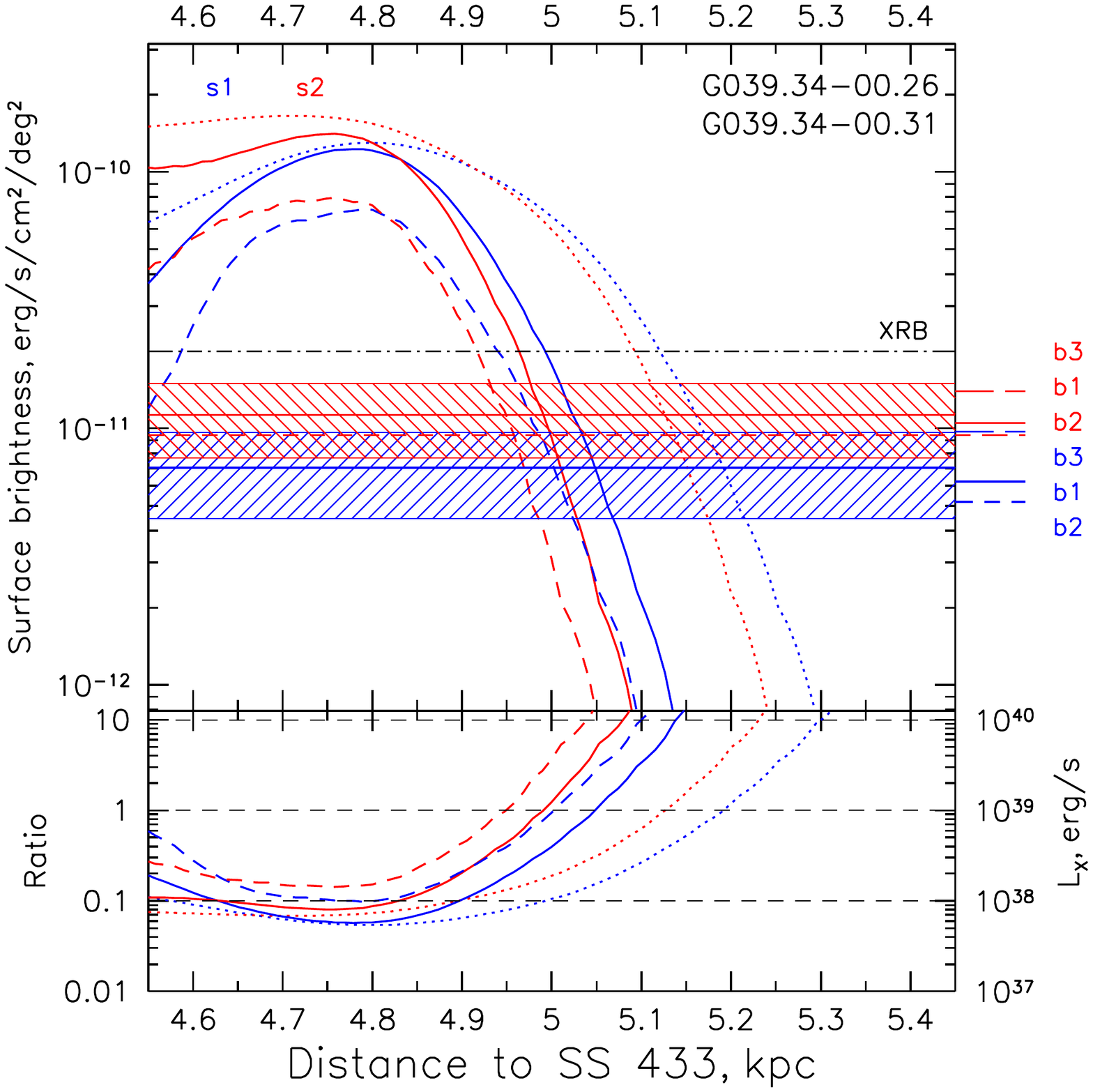}
\caption{\textit{Left panel.} Dependence of the predicted reflected emission surface brightness (for $L_X=10^{39}$ erg/s and $\Theta_r=50\deg$) on the line-of-sight location of the illuminating source. The solid blue and red lines correspond to source regions \textit{s1} and \textit{s2}, respectively (see Figure 4). The hatched regions show the measured surface brightness in these regions (blue-\textit{s1}, red-\textit{s2}) with the 1$\sigma$ uncertainty obtained by combining separate measurements based on background subtraction from three different regions. The levels of these individual measurements are marked on the right side (blue-\textit{s1}, red-\textit{s2}) with top-to-bottom ordering as given by the \textit{b1}-\textit{b3} labels. Clearly, the individual measurements are broadly consistent with the combined values. The black dash-dotted line marks the expected level of X-ray background in these regions. The bottom part shows the ratios of the predicted surface brightness to the measured one (left axis) and the derived upper limit on the apparent luminosity (right axis). \textit{Right panel.} Similar to the left panel, but with the predicted surface brightness dependences and corresponding ratios for $\Theta_r=21\deg=\Theta_p$ (solid lines) and $\Theta_r=10\deg$ (dashed lines) shown in comparison to $\Theta_r=50\deg$ (dotted lines). } 
\label{f:3c396constr}
\end{figure*}
%%%%%%%%%%%%%%%%%%%%%%%%%%%%%%%%%%%%%%%%%%%%%%%%%%%%%%%%%%%%%

%#######################################################
\subsection{G041.04-00.26}
%%%%%%%%%%%%%%%%%%%%%%%%%%%%%%%%%%%%%%%%%%%%%%%%%%%%%%%%%%%%%

\textit{Chandra} observation ObsId 1042 has been taken in ACIS-S configuration with a total exposure time of 66 ks, and its primary target was supernova remnant 3C 397 \citep{2005ApJ...618..321S}. Similar to the previous case,  we fully excluded the aim-point chip (S3) to avoid contamination by the extended emission of this supernova remnant. The resulting map of the 4-8 keV surface brightness is shown in Figure 6 (upper panel). No significant excess over the background level is visible. 

Compared to the previous case, the situation is much simpler, since only one cloud can potentially contribute to the reflection signal, and it is predicted to be not very extended (unfortunately, the brightest spot of the predicted emission falls onto the chip gap, see Figure 6). Because of that, we used single source and background regions. The expected surface brightness of the reflected emission inside the sources region is $\sim3\times10^{-11}$ erg/s/cm$^2$/deg$^2$ for $L_{X}=10^{39}$ erg/s and $\Theta_r=50\deg$ if the illuminating source is at $ d_{SS433}= 4.7$ kpc (see the middle panel in Figure 6). The selected background region is predicted to contain only a negligible contribution of the reflected emission (see the bottom panel in Figure 7). Because G041.04-00.26 is located significantly off-axis ($\approx 50\deg$) with respect to the SS 433 jets' precession cone, it can be illuminated only for $ \Theta_{r}\gtrsim30\deg $ \citep[see Figure 4 in][]{2016MNRAS.457.3963K}. Also, the fraction of time it is expected to spend inside the illumination cone (i.e. illumination `duty-cycle') is $\gtrsim 0.3$ for $ \Theta_r\gtrsim40\deg$, but then drops rapidly to zero for $ \Theta_r\lesssim 30\deg$ (see Figure 2 in \cite{2016MNRAS.457.3963K}). Hence, the constraints obtained for this cloud are relevant mostly for  large opening angles.

The measured background-subtracted 4-8 keV surface brightness is $(12\pm2)\times10^{-12}$ erg/s/cm$^2$/deg$^2$, and the upper limit on the reflected X-ray emission, $1.4\times10^{-11}$ erg/s/cm$^2$/deg$^2$, turns out to be quite similar to the limits obtained in the previous case. This results in an upper limit on the apparent luminosity of $\sim5\times 10^{38}$ erg/s for $ d_{SS433}\lesssim 4.8$ kpc and $\lesssim3\times 10^{39}$ erg/s for $ d_{SS433}=4.8-5$ kpc, assuming $\Theta_r=50\deg$ (see the bottom panel in Figure 6). Once again, due to the off-axis position of this cloud, the constraints for $\Theta_r=30\deg$ are significantly weaker,  $L_X\lesssim10^{39}$ erg/s for $ d_{SS433}\lesssim 4.7$ kpc, and $\lesssim 10^{40}$ erg/s up to $ d_{SS433}=4.85$ kpc.

The constraints derived from this cloud are noticeably weaker than those obtained in the case of G039.34-00.31 and G039.34-00.26. However, these are very important since they diminish the conditionality of the obtained results, i.e. their dependence on the incidence of particular clouds inside SS 433's illumination cone (taking into account the uncertainties in the distances to the individual clouds).  

%%%%%%%%%%%%%%%%%%%%%%%%%%%%%%%%%%%%%%%%%%%%%%%%%%%%%%%%%%%%%
  \subsection{G036.14+00.09}
%%%%%%%%%%%%%%%%%%%%%%%%%%%%%%%%%%%%%%%%%%%%%%%%%%%%%%%%%%%%%

\textit{Chandra} observation ObsId  12557 has been taken in ACIS-I configuration with a total exposure time of 38 ks, and its primary target was the young pulsar PSR J1856+0245 \citep{2012A&A...544A...3R}. The source is not very bright, $\lesssim10^{-13}$ erg/s/cm$^2$, and there are no signatures of extended emission associated with it at even lower level \citep{2012A&A...544A...3R}, so it can be readily excised by the standard procedure as for other point sources. This allows us to use almost the whole extent of the four chips for the source and background extraction regions. The resulting map of the 4-8 keV surface brightness is shown in Figure 7 (upper panel). No significant excess over the background level is visible here either. 

The cloud G036.14+00.09 is one of the least massive clouds in our sample (see Table 1), being an order of magnitude less massive than the three clouds considered above. As a result, the predicted X-ray reflection signal for it is well below the XRB level even for $L_X=10^{39}$ erg/s and $\Theta_{r}=50\deg$, reaching maximal values for $d_{SS433}\approx4.9$ kpc (see the middle panel in Figure 7). However, thanks to its larger projected distance from SS 433, it can fall inside the illumination cone for a larger range of $d_{SS433}$, as long as $\Theta_r$ is big enough to compensate for the relatively large off-axis angle ($\sim 35\deg$) for this cloud (see Figure 2 and Figure 4 in \cite{2016MNRAS.457.3963K}). For $\Theta_r<35\deg$, the probed range of $d_{SS433}$ becomes quite narrow with a center around 4.8 kpc.

The measured background-subtracted 4-8 keV surface brightness equals  $(11\pm5)\times10^{-12}$ erg/s/cm$^2$/deg$^2$, so that the upper limit on the reflected X-ray emission, $1.6\times10^{-11}$ erg/s/cm$^2$/deg$^2$, is once again similar to the limits set in the previous cases. For $\Theta_{r}=50\deg$, the resulting upper limit on the apparent luminosity is $\sim5\times 10^{39}$ erg/s over a very broad range of distances, $ d_{SS433}=4.7-5.2$ kpc, while it is below $10^{40}$ erg/s for all $ d_{SS433}<5.3$ kpc (see the bottom panel in Figure 7). For $\Theta_{r}=21\deg$, the $L_X<10^{40}$ erg/s constraint is limited to  $ d_{SS433}=4.65-4.95$ kpc, with the best limit ($L_X<3\times 10^{39}$ erg/s reached at $ d_{SS433}\approx4.8$ kpc) unchanged.

Thus, the main contribution of this cloud is that it provides a moderate upper limit over a very broad range of $d_{SS433}$ given that $\Theta_{r}\gtrsim 35\deg$, particularly for $d_{SS433}>5.1$ kpc where the constraints derived earlier were very poor (see Figures 5 and 6). For smaller $\Theta_r$, this cloud strengthens the independence of the obtained $L_X$ on the uncertainties in individual clouds positions for $d_{SS433}$ around 4.8 kpc.

%%%%%%%%%%%%%%%%%%%%%%%%%%%%%%%%%%%%%%%%%%%%%%%%%%%%%%%%%%%%%
\begin{figure}
\center
\includegraphics[width=0.95\columnwidth,bb=50 200 600 630]{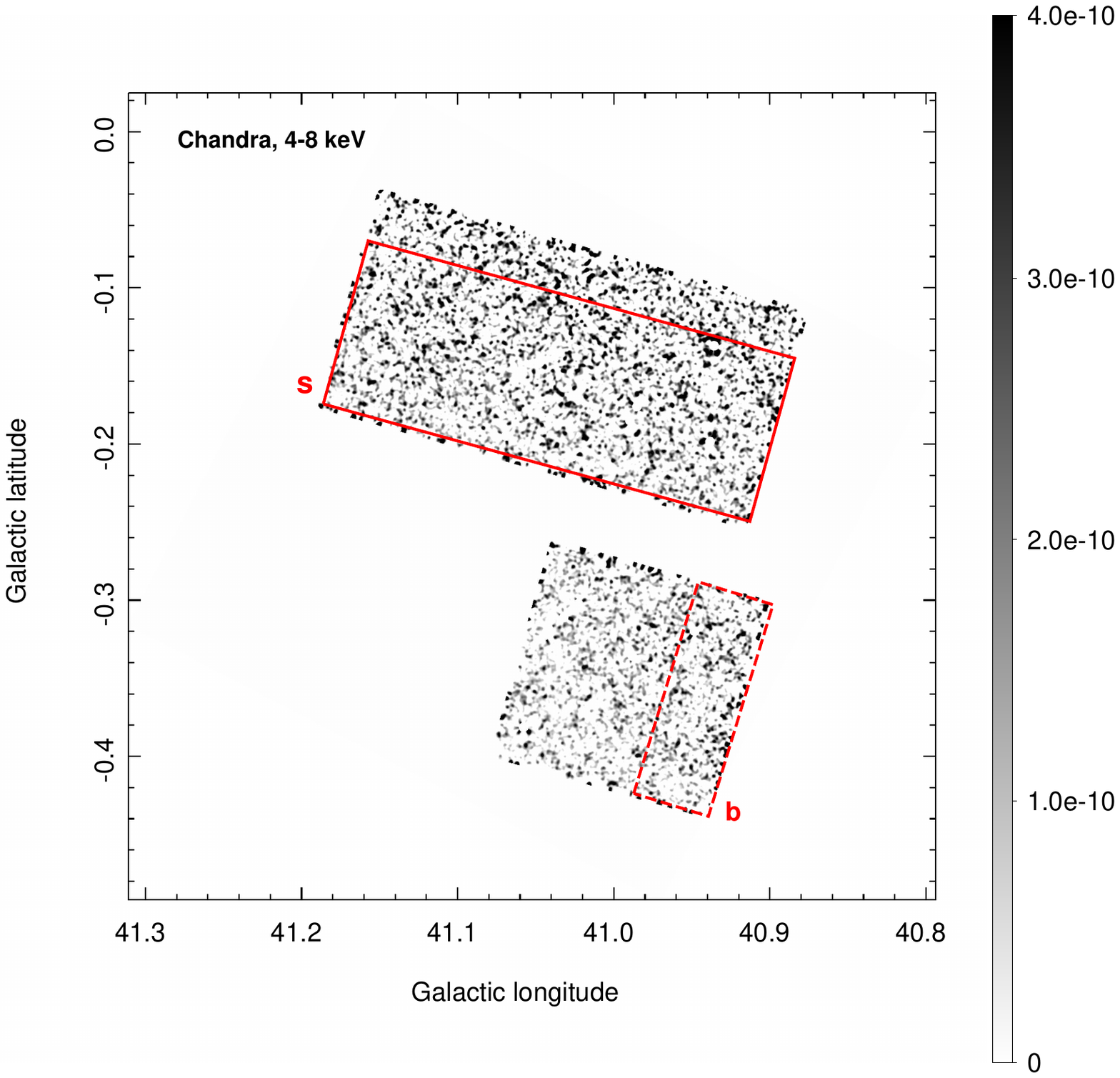}
\includegraphics[width=0.95\columnwidth,bb=50 200 600 650]{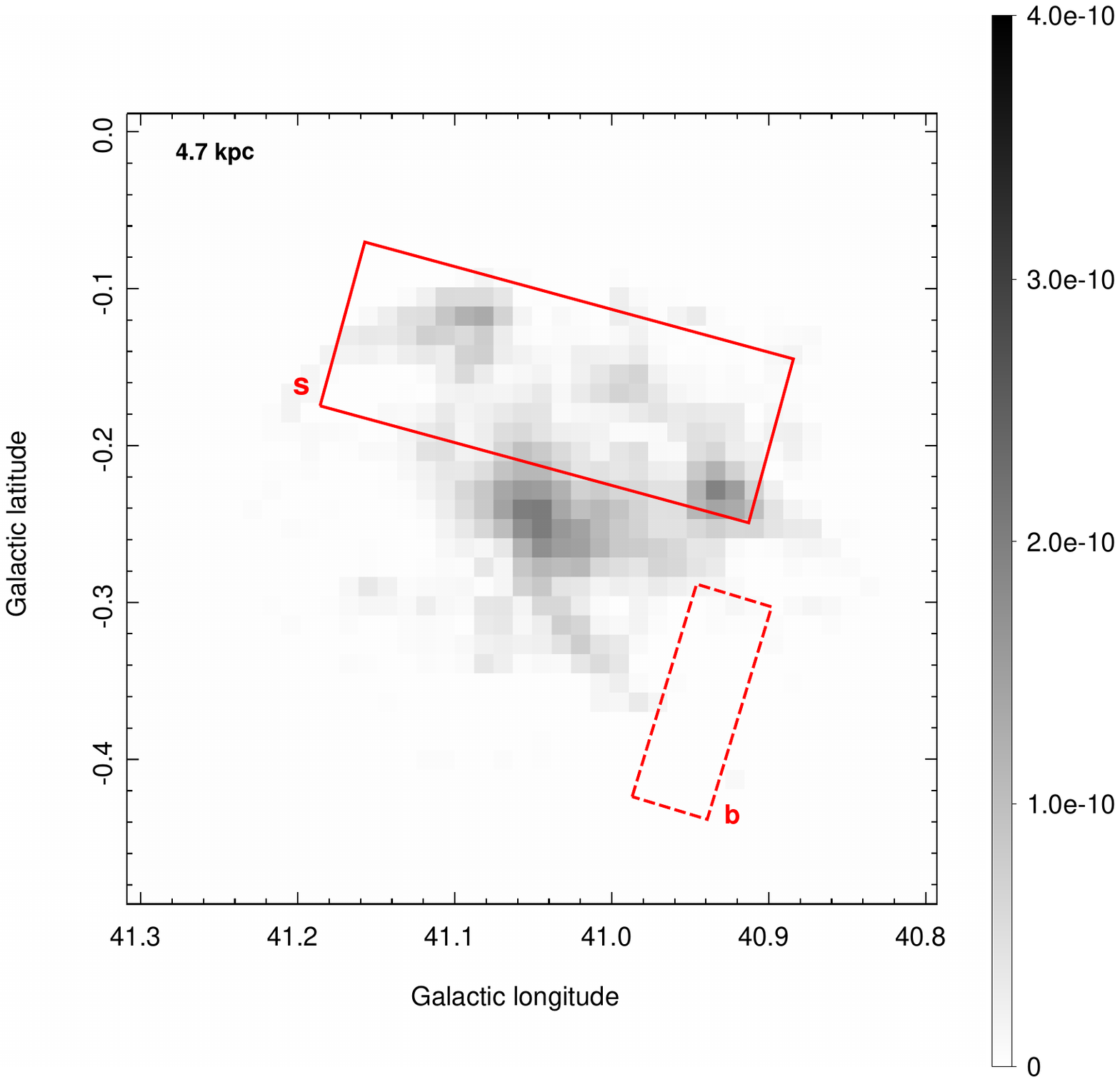}
\includegraphics[width=0.7\columnwidth,bb=100 180 550 730]{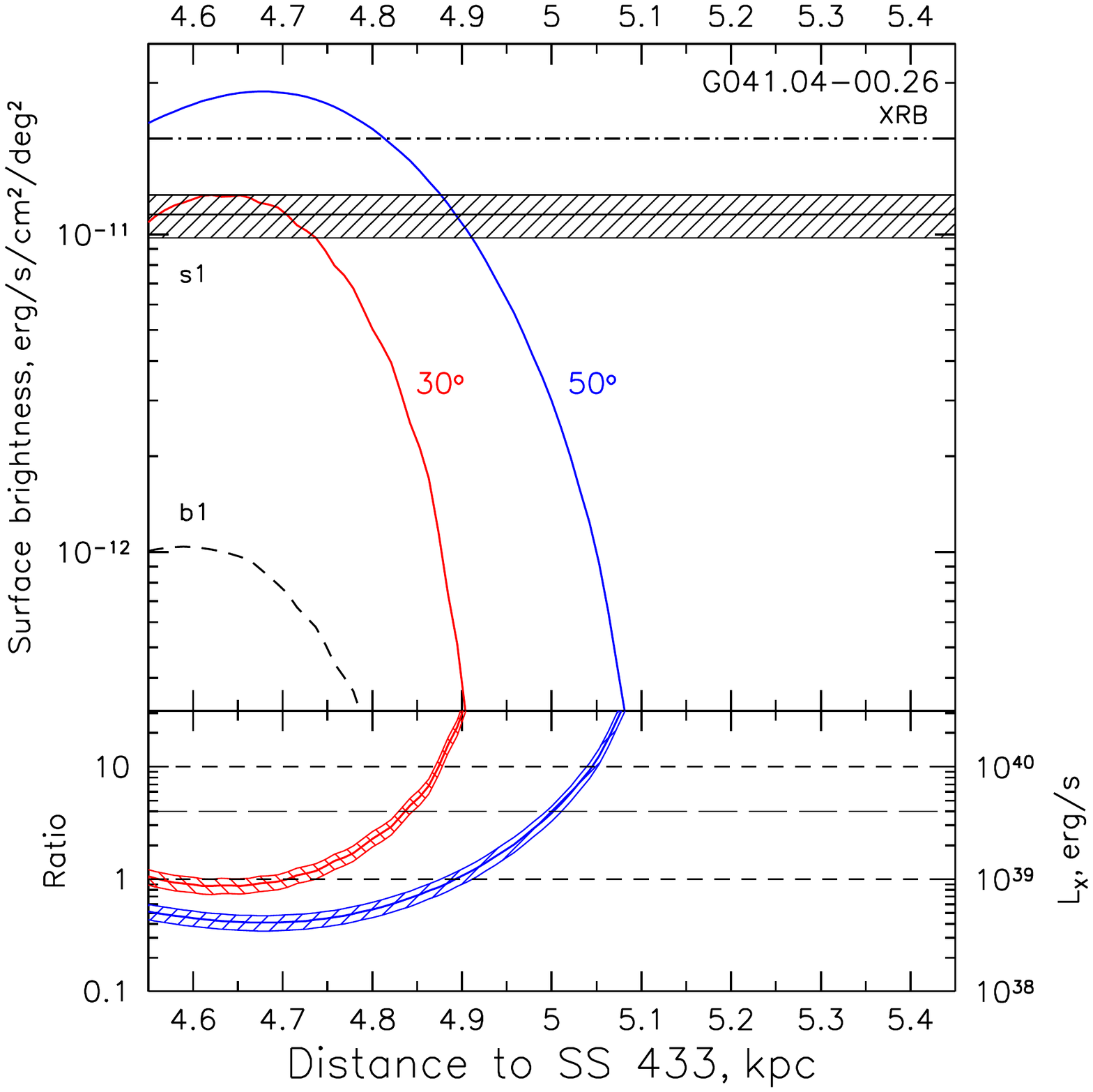}
\caption{Same as Figure 4, for the region where X-ray reflection on the cloud G041.04-00.26 is expected. The bottom panel shows the predicted surface brightness of the reflected emission inside the source region for $\Theta_r=50\deg $ (solid blue line) and $\Theta_r=30\deg $ (solid red line), and in the background region (dashed line), as functions of the assumed SS 433 distance. The hatched black region shows the measured surface brightness in the source region. In the bottom part of this panel, the ratios of the predicted to measured surface brightness are shown. The long-dashed line marks the $4\times 10^{39}$ erg/s level (see text).
} 
\label{f:3C397}
\end{figure}
%%%%%%%%%%%%%%%%%%%%%%%%%%%%%%%%%%%%%%%%%%%%%%%%%%%%%%%%%%%%%

%%%%%%%%%%%%%%%%%%%%%%%%%%%%%%%%%%%%%%%%%%%%%%%%%%%%%%%%%%%%%
\begin{figure}
\center
\includegraphics[width=0.95\columnwidth,bb=100 200 550 590]{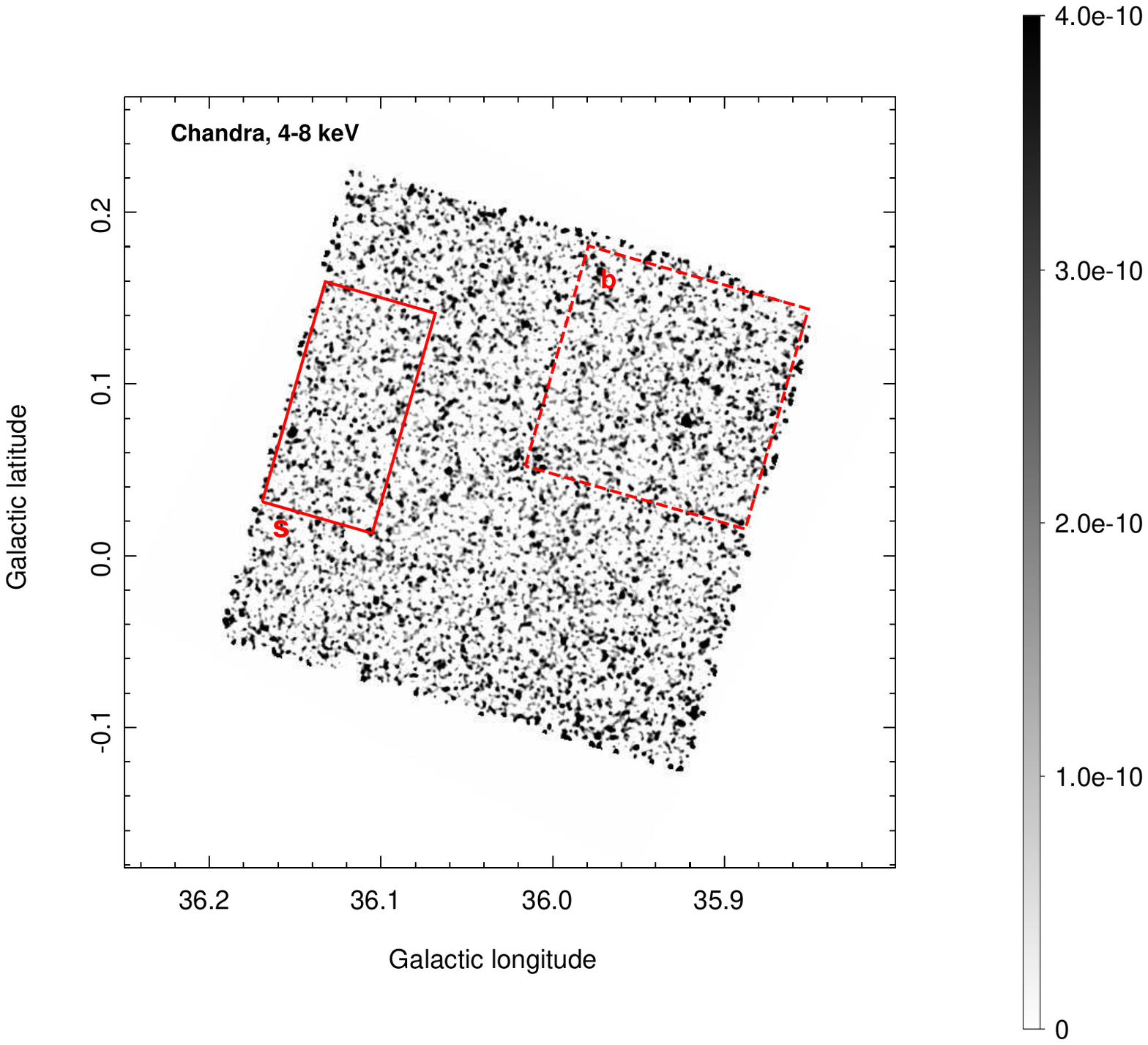}
\includegraphics[width=0.95\columnwidth,bb=100 200 550 620]{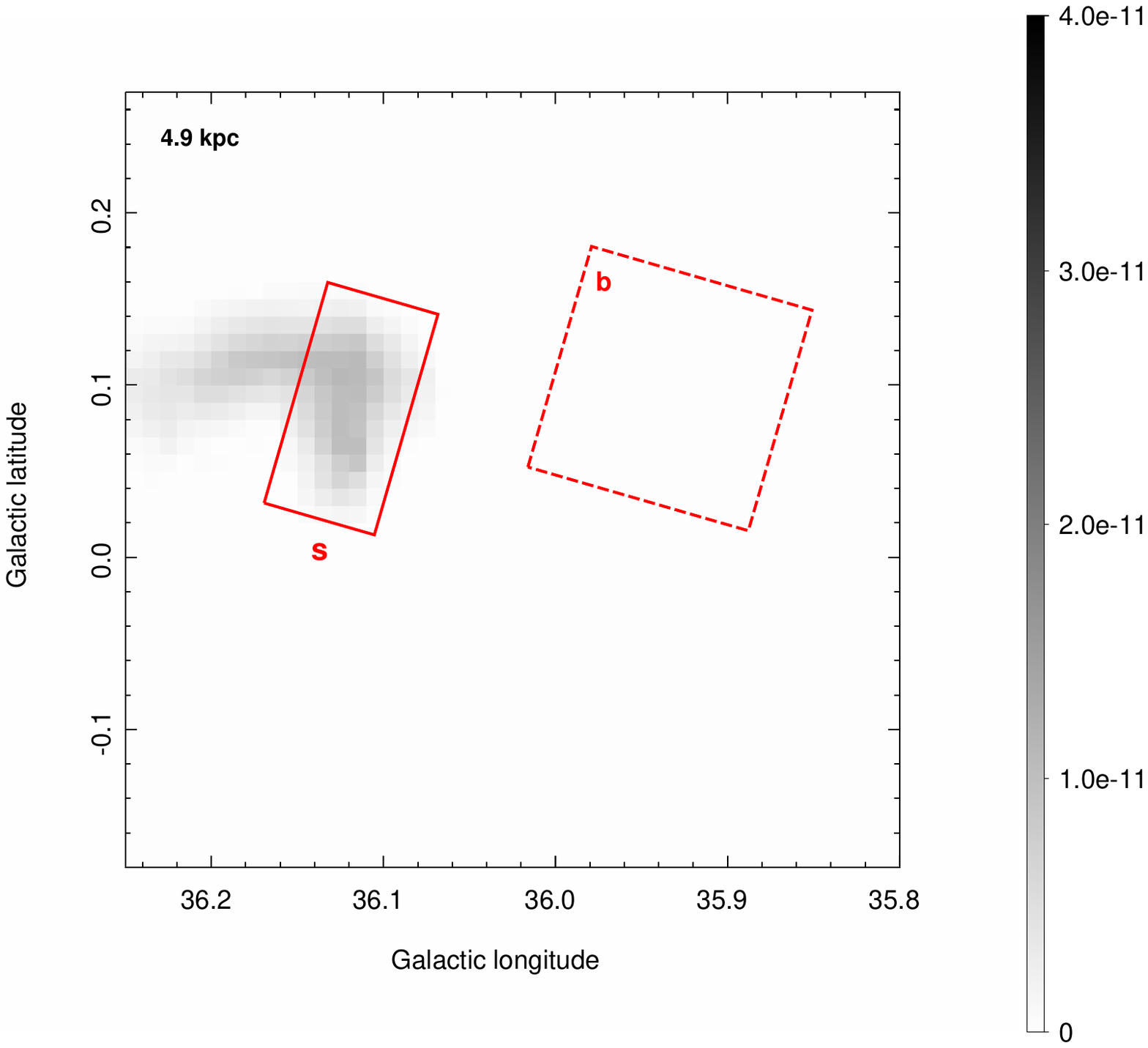}
\includegraphics[width=0.7\columnwidth,bb=100 180 550 730]{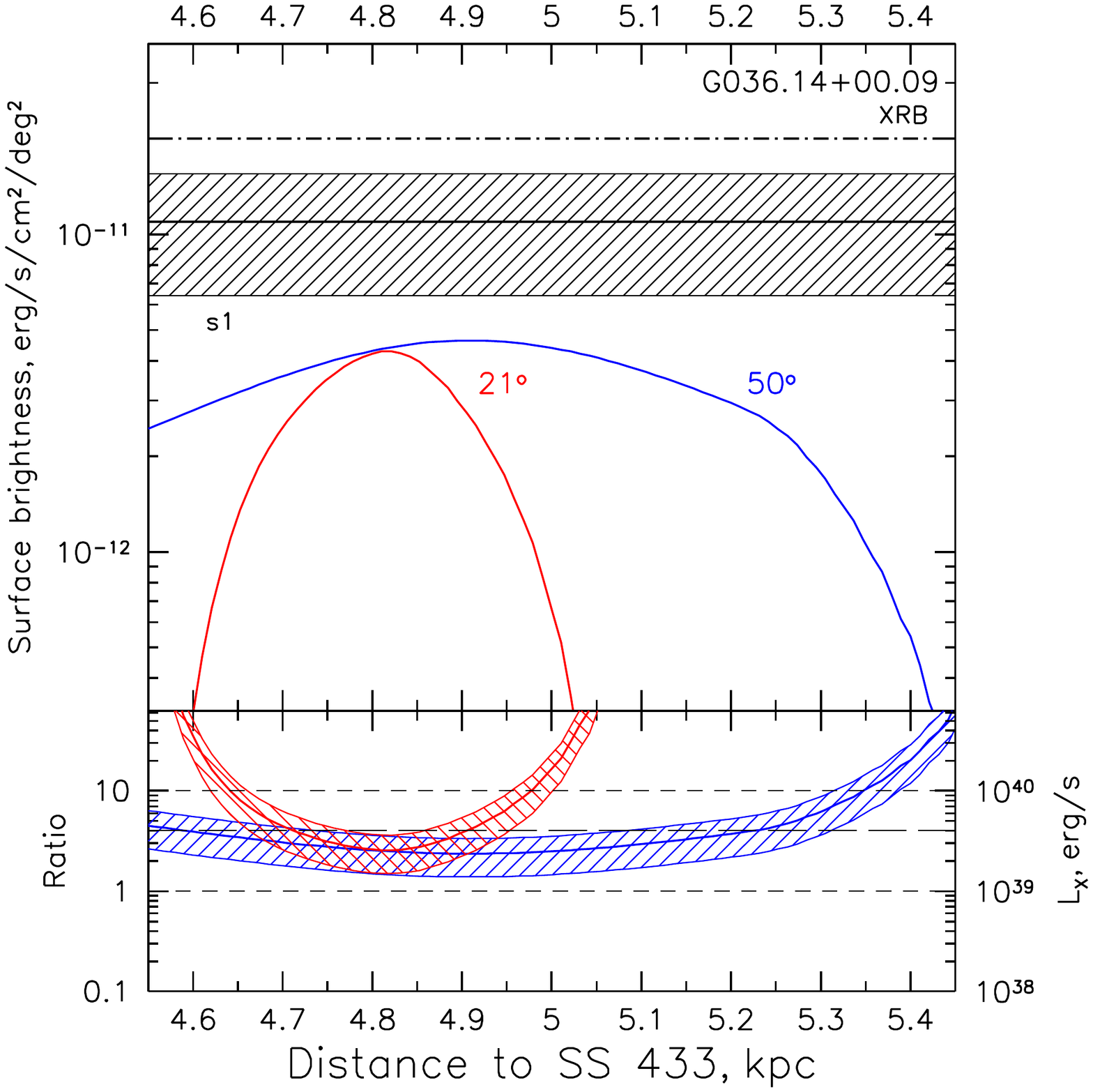}
\caption{Same as Figure 6, for the region where X-ray reflection on the cloud G036.14+00.09 is expected. Note the factor of 10 difference in the colorbar scale of the middle panel due to the significantly smaller mass of this cloud (see Table 1).
} 
\label{f:PSRJ1856}
\end{figure}
%%%%%%%%%%%%%%%%%%%%%%%%%%%%%%%%%%%%%%%%%%%%%%%%%%%%%%%%%%%%%

%%%%%%%%%%%%%%%%%%%%%%%%%%%%%%%%%
\section{5. Discussion and conclusions}
\label{s:discussion}
%%%%%%%%%%%%%%%%%%%%%%%%%%%%%%%%%

%%%
\begin{figure}
%origin: /home/ikh/Dropbox/ssrefl/paper/figs/3d/angles.svg
\includegraphics[width=\columnwidth,bb=30 180 600 700]{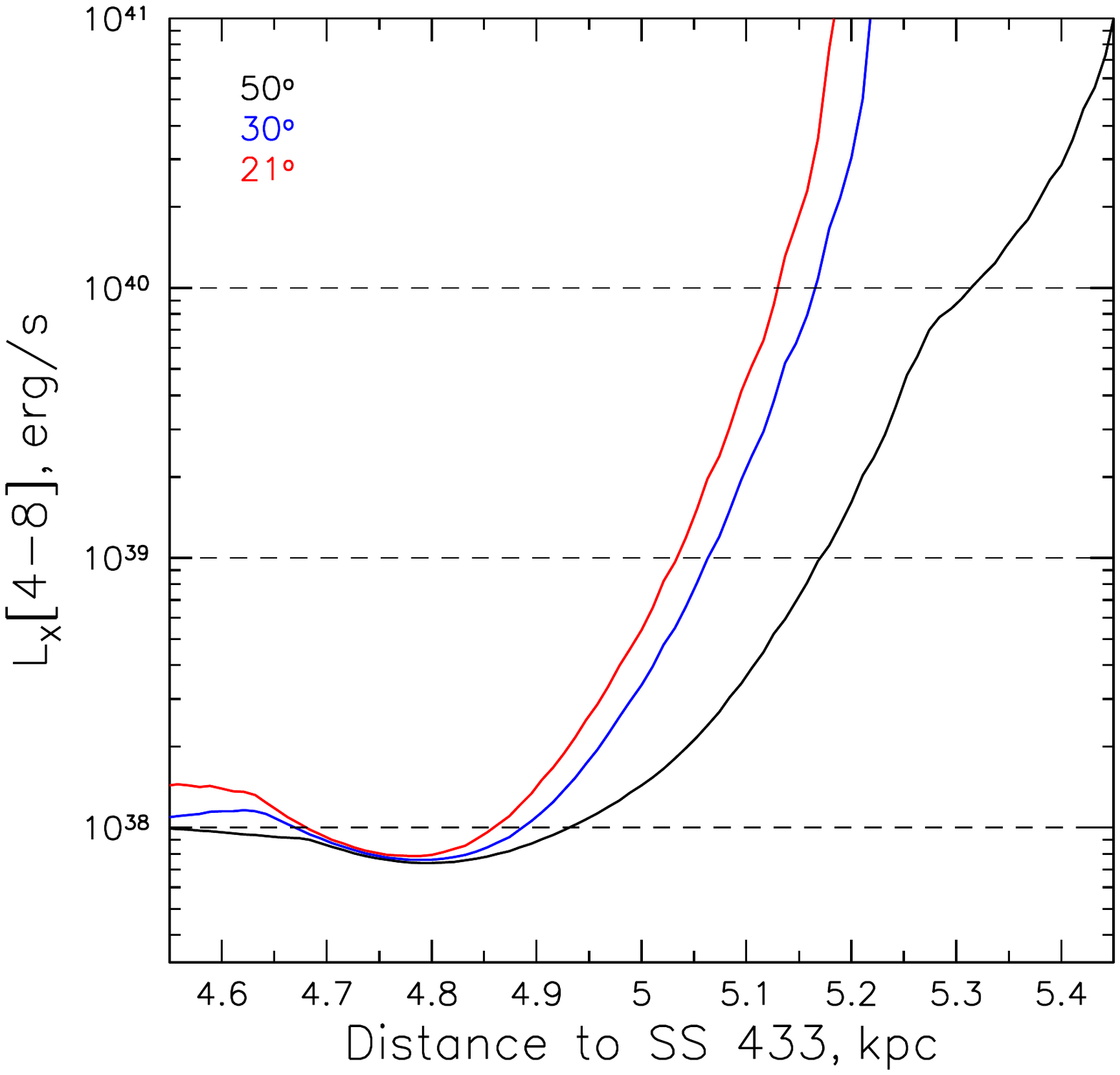}
\caption{Combined upper limits on the apparent 4-8 keV luminosity of SS 433 as a function of its line-of-sight position based on the constraints obtained for the individual clouds. The black curve corresponds to the half-opening angle of the collimation cone $\Theta_r=50\deg$, the blue line - $\Theta_r=30\deg$, the red line $\Theta_r=\Theta_p=21\deg$. } 
\label{f:combconstr}
\end{figure}
%%%%%%%%%%%%%%%%%%%%%%%%%%%%%%%%%%%%%%%%%%%%%%%%%%%%%%%%%%%%%

In the previous section, we obtained upper limits on the apparent 4-8 keV luminosity of SS 433 based on the constraints on the reflection signal from individual clouds. We can combine these upper limits, assuming that the distance estimates for all of the clouds are accurate. The resulting curves for different collimation angles are shown in Figure 8. As discussed before, the main effect is provided by the clouds G039.34-00.31 and G039.34-00.26, which are quite massive and located close to the projection of the jets' precession axis on the sky (see Figure 1). 

For the maximal collimation angle, $\Theta_r\approx50\deg$, the upper limit is $\sim 10^{38}$ erg/s for $d_{SS433}<5$ kpc and then grows rapidly to $\sim 10^{39}$ erg/s for $d_{SS433}\approx5.15$ kpc and $\sim 10^{40}$ erg/s for $d_{SS433}\approx5.3$ kpc. For  $d_{SS433}>5.3$ kpc, the considered clouds, unfortunately, do not provide significant constraints even for the largest collimation angle. For smaller collimation angles $\Theta_r\approx \Theta_p=21\deg$, the situation does not change strongly for  $d_{SS433}<4.9$ kpc, namely the upper limit remains as low as $\lesssim 2\times 10^{38}$ erg/s, while it grows to $\sim 5\times 10^{38}$ erg/s for $d_{SS433}\approx5$ kpc and $\sim 10^{40}$ erg/s for $d_{SS433}\approx5.15$ kpc. Thus, even for moderate collimation angles, the obtained constraint is very strong for a broad range of $d_{SS433}$.  

In reality, however, the uncertainties in distances to the individual clouds are quite large ($\gtrsim$ hundred parsecs, see Table 1), so that the actual disposition of the clouds with respect to SS 433 can significantly differ from that assumed in our modelling. One might then consider how many clouds do contribute to the constraint for each $ d_{SS433}$. As can be seen from Figures 5, 6 and 7, for $ d_{SS433}=4.75\pm0.1$ kpc and $\Theta_r\gtrsim30\deg$, each of the four considered clouds (located at $4.55\pm0.18$ kpc, $4.72\pm0.17$ kpc, $4.93\pm0.46$ kpc, and $5.15\pm0.25$ kpc, see Table 1) provides a constraint not worse than $\sim 4\times 10^{39}$ erg/s (this level is marked with the long-dashed horizontal line in the bottom panels of Figure 6 and 7). This, of course, significantly decreases the sensitivity of the obtained result to the actual incidence of individual clouds inside the illumination cone.

We do not consider specifically the $\Theta_r\ll \Theta_p=21\deg$ case (except for the curve for $\Theta_r=10\deg$ in Figure 5), because the size of the illumination region then becomes comparable to the typical size of the cloud and one also needs to take into account the duty cycle of illumination. As a result, the corresponding limits depend on $\Theta_r$ and other parameters in a complicated manner, not allowing us to draw any firm conclusions. For small collimation angles, it might be better to consider reflection on atomic gas, even though the expected surface brightness of the reflected emission is much lower than for molecular clouds. 

Finally, we can convert the obtained limits from the 4-8 keV energy range to the more commonly used 2-10 keV range. This can be done using the conversion factors presented in Appendix of \cite{2016MNRAS.457.3963K}. Namely, for a power-law spectrum with a high-energy cutoff, this factor varies from 2 to 4 if $\Gamma=2$ and the cutoff energy ranges between 3 and 10 keV. Hence,  for $\Theta_r\gtrsim21\deg$ and $d_{SS433}<4.9$ kpc, we get a limit $ L_{X,2-10}<8\times 10^{38}$ erg/s for the apparent 2-10 keV luminosity. Taken at face value, this result would effectively exclude association of SS 433 with ULXs. However, this result mainly comes from the constraint obtained from two clouds and there is a fair possibility that in reality they do not fall inside the illumination cone. The most conservative estimate, based on the constraints from the four clouds, {is $ L_{X,2-10}\lesssim 10^{40}$ erg/s if $ d_{SS433}=4.65-4.85$ kpc. Thus, we conclude that SS 433 is not likely to belong to the brightest ultraluminous X-ray sources if it could be observed face-on (unless its X-ray emission is strongly collimated), despite the extremely high mass transfer rate in the system. Clearly, better X-ray coverage of the molecular clouds in the region of interest is needed to extend applicability of this result for a larger range of $d_{SS433}$ and eliminate its dependence on incidence of individual clouds inside the putative illumination cone.}

{This result might also indicate that SS 433 is rather} an ultraluminous supersoft source \citep{2016MNRAS.456.1859U,2016MNRAS.457.3963K}. Such sources have apparent X-ray luminosities of few$\times 10^{39}$ erg/s with virtually no emission above 2 keV. Of course, the considered reflection signal does not allow us to constrain the source's luminosity below 2 keV, since the reflection albedo (essentially, the ratio of the scattering and photo-absorption cross-sections) at these energies is very small. However, in this case one might explore the possible impact of such emission on the closer environment of SS 433 \citep{1993AstL...19...41P,2016MNRAS.457.3963K,2017AstL...43..388K,2018arXiv181112564W}. 

Finally, we note that the obtained constraints correspond to the luminosity averaged over significant time intervals due to light-travel-time effects. Namely, the light-crossing time of the typical cloud amounts to $\sim50$ years, while the propagation of X-rays from SS 433 to any of the considered clouds takes $ \sim 600 $ years or more. SS 433 is observed to be very stable over the past 40 years, for which observations are available \citep[e.g.][]{2018ARep...62..747C}, while the structures observed inside the W 50 nebula imply that the jets have been active for more than $\sim$1000 years \citep{2011MNRAS.414.2838G,2017A&A...599A..77P}. Also, the latter timescale is shorter then any evolutionary time scales of the system \citep[e.g.][]{2017MNRAS.465.2092P,2017MNRAS.471.4256V}. Hence, the obtained constraints may be regarded as limits on SS 433 present-day apparent luminosity. 

%
%%%%%%%%%%%%%%%%%%%%%%%%%%%%%%%%%
%	The acknowledgements
%%%%%%%%%%%%%%%%%%%%%%%%%%%%%%%%%
\section{Acknowledgements}
The research was supported by the Russian Science Foundation (grant 14-12-01315). We are sincerely grateful to Eugene Churazov for providing us with the software package for preparation, reduction, and analysis of \textit{Chandra} data, help with it and valuable discussions.
%%%%%%%%%%%%%%%%%%%%%%%%%%%%%%%%%
%	The bibliography
%%%%%%%%%%%%%%%%%%%%%%%%%%%%%%%%%

%%%%%%%%%%%%%%%%%%%%%%%%%%%%%%%%%
%	The end
%%%%%%%%%%%%%%%%%%%%%%%%%%%%%%%%%

\begin{thebibliography}{}
%
\bibitem[\protect\citeauthoryear{Abramowicz, Calvani, \& Nobili}{1980}]{1980ApJ...242..772A} Abramowicz M.~A., Calvani M., Nobili L., 1980, ApJ, 242, 772 

\bibitem[\protect\citeauthoryear{Bachetti et al.}{2014}]{2014Natur.514..202B} Bachetti M., et al., 2014, Natur, 514, 202 

\bibitem[\protect\citeauthoryear{Begelman, King, \& Pringle}{2006}]{2006MNRAS.370..399B} Begelman M.~C., King A.~R., Pringle J.~E., 2006, MNRAS, 370, 399 

\bibitem[\protect\citeauthoryear{Blundell \& Bowler}{2004}]{2004ApJ...616L.159B} Blundell K.~M., Bowler M.~G., 2004, ApJ, 616, L159 


\bibitem[\protect\citeauthoryear{Brinkmann, Kotani, \& Kawai}{2005}]{2005A&A...431..575B} Brinkmann W., Kotani T., Kawai N., 2005, A\&A, 431, 575 

\bibitem[\protect\citeauthoryear{Caballero-Garcia et al.}{2018}]{2018arXiv180207149C} Caballero-Garcia M.~D., Fabrika S., Castro-Tirado A.~J., Bursa M., Dovciak M., Castellon A., Karas V., 2018, arXiv, arXiv:1802.07149 

\bibitem[\protect\citeauthoryear{Carpano et al.}{2018}]{2018MNRAS.476L..45C} Carpano S., Haberl F., Maitra C., Vasilopoulos G., 2018, MNRAS, 476, L45 

\bibitem[\protect\citeauthoryear{Cherepashchuk et al.}{2018}]{2018ARep...62..747C} Cherepashchuk A.~M., Esipov V.~F., Dodin A.~V., Davydov V.~V., Belinskii A.~A., 2018, ARep, 62, 747 

\bibitem[\protect\citeauthoryear{Cherepashchuk, Postnov, \& Belinski}{2018}]{2018MNRAS.479.4844C} Cherepashchuk A.~M., Postnov K.~A., Belinski A.~A., 2018, MNRAS, 479, 4844 

\bibitem[\protect\citeauthoryear{Churazov, Sunyaev, \& Sazonov}{2002}]{2002MNRAS.330..817C} Churazov E., Sunyaev R., Sazonov S., 2002, MNRAS, 330, 817 

\bibitem[\protect\citeauthoryear{Churazov et al.}{2012}]{2012MNRAS.421.1123C} Churazov E., et al., 2012, MNRAS, 421, 1123 


\bibitem[\protect\citeauthoryear{Churazov et al.}{2017a}]{2017MNRAS.465...45C} Churazov E., Khabibullin I., Sunyaev R., Ponti G., 2017a, MNRAS, 465, 45 

\bibitem[\protect\citeauthoryear{Churazov et al.}{2017b}]{2017MNRAS.468..165C} Churazov E., Khabibullin I., Ponti G., Sunyaev R., 2017b, MNRAS, 468, 165 

\bibitem[\protect\citeauthoryear{Churazov et al.}{2017c}]{2017MNRAS.471.3293C} Churazov E., Khabibullin I., Sunyaev R., Ponti G., 2017c, MNRAS, 471, 3293 

\bibitem[\protect\citeauthoryear{Dolan et al.}{1997}]{1997A&A...327..648D} Dolan J.~F., et al., 1997, A\&A, 327, 648 

\bibitem[\protect\citeauthoryear{Fabrika \& Mescheryakov}{2001}]{Fabrika2001} Fabrika S., Mescheryakov A., 2001, IAUS, 205, 268 

\bibitem[\protect\citeauthoryear{Fabrika}{2004}]{2004ASPRv..12....1F} Fabrika S., 2004, ASPRv, 12, 1 

\bibitem[\protect\citeauthoryear{Fabrika et al.}{2015}]{2015NatPh..11..551F} Fabrika S., Ueda Y., Vinokurov A., Sholukhova O., Shidatsu M., 2015, NatPh, 11, 551 

\bibitem[\protect\citeauthoryear{F{\"u}rst et al.}{2016}]{2016ApJ...831L..14F} F{\"u}rst F., et al., 2016, ApJ, 831, L14 

\bibitem[\protect\citeauthoryear{Gilfanov, Grimm, \& Sunyaev}{2004}]{2004NuPhS.132..369G} Gilfanov M., Grimm H.-J., Sunyaev R., 2004, NuPhS, 132, 369 

\bibitem[\protect\citeauthoryear{Goodall, Alouani-Bibi, \& Blundell}{2011}]{2011MNRAS.414.2838G} Goodall P.~T., Alouani-Bibi F., Blundell K.~M., 2011, MNRAS, 414, 2838 

\bibitem[\protect\citeauthoryear{Israel et al.}{2017}]{2017MNRAS.466L..48I} Israel G.~L., et al., 2017, MNRAS, 466, L48 

\bibitem[\protect\citeauthoryear{Kaaret, Feng, \& Roberts}{2017}]{2017ARA&A..55..303K} Kaaret P., Feng H., Roberts T.~P., 2017, ARA\&A, 55, 303 
\bibitem[\protect\citeauthoryear{Kawashima et al.}{2012}]{2012ApJ...752...18K} Kawashima T., Ohsuga K., Mineshige S., Yoshida T., Heinzeller D., Matsumoto R., 2012, ApJ, 752, 18 


\bibitem[\protect\citeauthoryear{Kalberla \& Kerp}{2009}]{2009ARA&A..47...27K} Kalberla P.~M.~W., Kerp J., 2009, ARA\&A, 47, 27 

\bibitem[\protect\citeauthoryear{King et al.}{2001}]{2001ApJ...552L.109K} King A.~R., Davies M.~B., Ward M.~J., Fabbiano G., Elvis M., 2001, ApJ, 552, L109 

\bibitem[\protect\citeauthoryear{Khabibullin, Medvedev, \& Sazonov}{2016}]{2016MNRAS.455.1414K} Khabibullin I., Medvedev P., Sazonov S., 2016, MNRAS, 455, 1414 

\bibitem[\protect\citeauthoryear{Khabibullin \& Sazonov}{2016}]{2016MNRAS.457.3963K} Khabibullin I., Sazonov S., 2016, MNRAS, 457, 3963 

\bibitem[\protect\citeauthoryear{Khabibullin \& Sazonov}{2017}]{2017AstL...43..388K} Khabibullin I.~I., Sazonov S.~Y., 2017, AstL, 43, 388 

\bibitem[\protect\citeauthoryear{Kotani et al.}{1996}]{1996PASJ...48..619K} Kotani T., Kawai N., Matsuoka M., Brinkmann W., 1996, PASJ, 48, 619 

\bibitem[\protect\citeauthoryear{Marshall, Canizares, \& Schulz}{2002}]{2002ApJ...564..941M} Marshall H.~L., Canizares C.~R., Schulz N.~S., 2002, ApJ, 564, 941 

\bibitem[\protect\citeauthoryear{Marshall et al.}{2013}]{2013ApJ...775...75M} Marshall H.~L., Canizares C.~R., Hillwig T., Mioduszewski A., Rupen M., Schulz N.~S., Nowak M., Heinz S., 2013, ApJ, 775, 75 

\bibitem[\protect\citeauthoryear{McKee \& Ostriker}{2007}]{2007ARA&A..45..565M} McKee C.~F., Ostriker E.~C., 2007, ARA\&A, 45, 565 

\bibitem[\protect\citeauthoryear{Medvedev \& Fabrika}{2010}]{2010MNRAS.402..479M} Medvedev A., Fabrika S., 2010, MNRAS, 402, 479 

\bibitem[\protect\citeauthoryear{Medvedev et al.}{2018}]{2018AstL...44..390M} Medvedev P.~S., Khabibullin I.~I., Sazonov S.~Y., Churazov E.~M., Tsygankov S.~S., 2018, AstL, 44, 390 

\bibitem[\protect\citeauthoryear{Migliari, Fender, \& M{\'e}ndez}{2002}]{2002Sci...297.1673M} Migliari S., Fender R., M{\'e}ndez M., 2002, Sci, 297, 1673 

\bibitem[\protect\citeauthoryear{Miller-Jones et al.}{2008}]{2008ApJ...682.1141M} Miller-Jones J.~C.~A., Migliari S., Fender R.~P., Thompson T.~W.~J., van der Klis M., M{\'e}ndez M., 2008, ApJ, 682, 1141 

\bibitem[\protect\citeauthoryear{Mineo, Gilfanov, \& Sunyaev}{2012}]{2012MNRAS.419.2095M} Mineo S., Gilfanov M., Sunyaev R., 2012, MNRAS, 419, 2095 

\bibitem[\protect\citeauthoryear{Molaro, Khatri, \& Sunyaev}{2014}]{2014A&A...564A.107M} Molaro M., Khatri R., Sunyaev R.~A., 2014, A\&A, 564, A107 


\bibitem[\protect\citeauthoryear{Ohsuga \& Mineshige}{2014}]{2014SSRv..183..353O} Ohsuga K., Mineshige S., 2014, SSRv, 183, 353 

\bibitem[\protect\citeauthoryear{Olbert et al.}{2003}]{2003ApJ...592L..45O} Olbert C.~M., Keohane J.~W., Arnaud K.~A., Dyer K.~K., Reynolds S.~P., Safi-Harb S., 2003, ApJ, 592, L45 

\bibitem[\protect\citeauthoryear{Panferov \& Fabrika}{1993}]{1993AstL...19...41P} Panferov A.~A., Fabrika S.~N., 1993, AstL, 19, 41 

\bibitem[\protect\citeauthoryear{Panferov}{2014}]{2014A&A...562A.130P} Panferov A., 2014, A\&A, 562, A130 

\bibitem[\protect\citeauthoryear{Panferov}{2017}]{2017A&A...599A..77P} Panferov A.~A., 2017, A\&A, 599, A77 

\bibitem[\protect\citeauthoryear{Pavlovskii et al.}{2017}]{2017MNRAS.465.2092P} Pavlovskii K., Ivanova N., Belczynski K., Van K.~X., 2017, MNRAS, 465, 2092 

\bibitem[\protect\citeauthoryear{Poutanen et al.}{2007}]{2007MNRAS.377.1187P} Poutanen J., Lipunova G., Fabrika S., Butkevich A.~G., Abolmasov P., 2007, MNRAS, 377, 1187 

\bibitem[\protect\citeauthoryear{Rappaport, Podsiadlowski, \& Pfahl}{2005}]{2005MNRAS.356..401R} Rappaport S.~A., Podsiadlowski P., Pfahl E., 2005, MNRAS, 356, 401 

\bibitem[\protect\citeauthoryear{Rathborne et al.}{2009}]{2009ApJS..182..131R} Rathborne J.~M., Johnson A.~M., Jackson J.~M., Shah R.~Y., Simon R., 2009, ApJS, 182, 131 

\bibitem[\protect\citeauthoryear{Revnivtsev et al.}{2006}]{2006A&A...452..169R} Revnivtsev M., Sazonov S., Gilfanov M., Churazov E., Sunyaev R., 2006, A\&A, 452, 169 

\bibitem[\protect\citeauthoryear{Roman-Duval et al.}{2009}]{2009ApJ...699.1153R} Roman-Duval J., Jackson J.~M., Heyer M., Johnson A., Rathborne J., Shah R., Simon R., 2009, ApJ, 699, 1153 


\bibitem[\protect\citeauthoryear{Roman-Duval et al.}{2010}]{2010ApJ...723..492R} Roman-Duval J., Jackson J.~M., Heyer M., Rathborne J., Simon R., 2010, ApJ, 723, 492 


\bibitem[\protect\citeauthoryear{Rousseau, et al.}{2012}]{2012A&A...544A...3R} Rousseau R., et al., 2012, A\&A, 544, A3

\bibitem[\protect\citeauthoryear{Safi-Harb, Dubner, Petre, Holt \& Durouchoux}{2005}]{2005ApJ...618..321S} Safi-Harb S., Dubner G., Petre R., Holt S.~S., Durouchoux P., 2005, ApJ, 618, 321


\bibitem[\protect\citeauthoryear{Sazonov \& Khabibullin}{2017a}]{2017MNRAS.466.1019S} Sazonov S., Khabibullin I., 2017a, MNRAS, 466, 1019 

\bibitem[\protect\citeauthoryear{Sazonov \& Khabibullin}{2017b}]{2017AstL...43..211S} Sazonov S.~Y., Khabibullin I.~I., 2017b, AstL, 43, 211 


\bibitem[\protect\citeauthoryear{Shakura \& Sunyaev}{1973}]{1973A&A....24..337S} Shakura N.~I., Sunyaev R.~A., 1973, A\&A, 24, 337 


\bibitem[\protect\citeauthoryear{Sugizaki et al.}{2001}]{2001ApJS..134...77S} Sugizaki M., Mitsuda K., Kaneda H., Matsuzaki K., Yamauchi S., Koyama K., 2001, ApJS, 134, 77 

\bibitem[\protect\citeauthoryear{Sunyaev \& Churazov}{1996}]{1996AstL...22..648S} Sunyaev R.~A., Churazov E.~M., 1996, AstL, 22, 648 

\bibitem[\protect\citeauthoryear{Urquhart \& Soria}{2016}]{2016MNRAS.456.1859U} Urquhart R., Soria R., 2016, MNRAS, 456, 1859 

\bibitem[\protect\citeauthoryear{van den Heuvel, Portegies Zwart, \& de Mink}{2017}]{2017MNRAS.471.4256V} van den Heuvel E.~P.~J., Portegies Zwart S.~F., de Mink S.~E., 2017, MNRAS, 471, 4256 

\bibitem[\protect\citeauthoryear{Vikhlinin et al.}{2005}]{2005ApJ...628..655V} Vikhlinin A., Markevitch M., Murray S.~S., Jones C., Forman W., Van Speybroeck L., 2005, ApJ, 628, 655 

\bibitem[\protect\citeauthoryear{Waisberg et al.}{2018}]{2018arXiv181112564W} Waisberg I., Dexter J., Olivier-Petrucci P., Dubus G., Perraut K., 2018, arXiv, arXiv:1811.12564 

\bibitem[\protect\citeauthoryear{Walton et al.}{2014}]{2014ApJ...793...21W} Walton D.~J., et al., 2014, ApJ, 793, 21 


%
\end{thebibliography}
\end{document}